\documentclass[conference]{IEEEtran}
\IEEEoverridecommandlockouts
\usepackage{cite}
\usepackage{amsmath,amssymb,amsfonts}
\usepackage{algorithmic}
\usepackage{graphicx}
\usepackage{textcomp}
\usepackage{xcolor}
\def\BibTeX{{\rm B\kern-.05em{\sc i\kern-.025em b}\kern-.08em
    T\kern-.1667em\lower.7ex\hbox{E}\kern-.125emX}}

\usepackage{soul}
\usepackage{url}
\usepackage{comment}
\usepackage{multirow}
\usepackage[ruled,linesnumbered]{algorithm2e}

\usepackage{floatrow}
\newcommand{\squeezeup}{\vspace{-2.5mm}}
\usepackage{subcaption} %multiple images on the same page

\renewcommand{\baselinestretch}{0.986}

\begin{document}

\newcommand{\name}{BWAP\xspace}
\newcommand{\bwapA}{canonical tuner\xspace}
\newcommand{\bwapB}{DWP tuner\xspace}
\newcommand{\DWP}{\emph{DWP}\xspace}
\newcommand{\uniformworkers}{\emph{uni\-form-workers}\xspace}
\newcommand{\uniformall}{\emph{uniform-all}\xspace}

\onecolumn

© 2020 IEEE.  Personal use of this material is permitted.  Permission from IEEE must be obtained for all other uses, in any current or future media, including reprinting/republishing this material for advertising or promotional purposes, creating new collective works, for resale or redistribution to servers or lists, or reuse of any copyrighted component of this work in other works.\\

If you'd like to cite this work, please use the reference below:\\

\noindent\emph{\textbf{Bandwidth-Aware Page Placement in NUMA Systems.\\
D. Gureya, J. Neto, R. Karimi, J. Barreto, P. Bhatotia, V. Quema, R. Rodrigues, P. Romano, and V. Vlassov.\\
34th IEEE International Parallel \& Distributed Processing Symposium (IPDPS), 2020.}}

\twocolumn

\newpage

\title{Bandwidth-Aware Page Placement in NUMA
%\thanks{This work was supported by Portuguese funds through Funda\c{c}\~{a}o para a Ci\^{e}ncia e Tecnologia via projects UID/CEC/50021/2013.}
}

\begin{comment}
\author{\IEEEauthorblockN{%1\textsuperscript{st} 
David Gureya, João Neto, Paolo Romano, Rodrigo Rodrigues, João Barreto}
\IEEEauthorblockA{INESC-ID, Instituto Superior Técnico, University of Lisbon, Portugal}
\and
\IEEEauthorblockN{Vivien Quema}
\IEEEauthorblockA{Grenoble INP/ENSIMAG, France}
\and
\IEEEauthorblockN{Pramod Bhatotia}
\IEEEauthorblockA{University of Edinburgh, UK}
\and
\IEEEauthorblockN{Reza Karimi}
\IEEEauthorblockA{Emory University, USA}
\and
\IEEEauthorblockN{Vladimir Vlassov}
\IEEEauthorblockA{KTH Royal Institute of Technology, Sweden}
}
\end{comment}
\author{\IEEEauthorblockN{David Gureya\IEEEauthorrefmark{1}\IEEEauthorrefmark{2},
João Neto\IEEEauthorrefmark{1},  
Reza Karimi\IEEEauthorrefmark{3},  
João Barreto\IEEEauthorrefmark{1},
Pramod Bhatotia\IEEEauthorrefmark{4},
Vivien Quema\IEEEauthorrefmark{5},
Rodrigo Rodrigues\IEEEauthorrefmark{1},\\
Paolo Romano\IEEEauthorrefmark{1}, and 
Vladimir Vlassov\IEEEauthorrefmark{2}}\\

\IEEEauthorblockA{\IEEEauthorrefmark{1}INESC-ID, Instituto Superior Técnico, University of Lisbon, Lisbon, Portugal\\
%Email: \{david.gureya,joaomlneto\}@tecnico.ulisboa.pt
}
\IEEEauthorblockA{\IEEEauthorrefmark{2} KTH Royal Institute of Technology, Stockholm, Sweden\\
%Email: \{daharewa,vladv\}@kth.se
}
\IEEEauthorblockA{\IEEEauthorrefmark{3} Emory University, Atlanta, GA, USA\\
%Email: rkarimi@emory.edu
}
\IEEEauthorblockA{\IEEEauthorrefmark{4} University of Edinburgh, Edinburgh, United Kingdom\\
%Email: xxxx
}
\IEEEauthorblockA{\IEEEauthorrefmark{5} Grenoble INP/ENSIMAG, Grenoble, France\\
%Email: xxxx
}
}

\maketitle

\pagestyle{plain}

\begin{abstract}
Page placement is a critical problem %to
for me\-mo\-ry-intensive applications running on a shared-memory multiprocessor with a non-uniform memory access (NU\-MA) architecture. %architectures.
%However,
State-of-the-art page placement mechanisms interleave pages evenly across NUMA nodes. However, this approach fails
to maximize memory throughput in modern NUMA systems, characterized by asymmetric bandwidths and latencies, and sensitive to memory contention and interconnect congestion phenomena.

We propose \name, a novel page placement mechanism 
%, unlike the state-of-the-art alternatives, 
based on %the principle of 
asymmetric weighted page interleaving.
\name combines an %application-agnostic, but architecture-aware 
analytical performance model of the target NUMA system with on-line iterative tuning of page distribution for a given memory-intensive application. %This allows \name to reach closer to the optimal page placement for the target application at runtime, and, as a consequence to improve parallel execution of the application.
Our experimental evaluation with %a representative set of
representative me\-mo\-ry-intensive workloads shows that \name performs up to $66\%$ better than state-of-the-art techniques. These gains are particularly relevant %This advantage is especially prevalent 
when multiple co-located applications run in disjoint partitions of a large NUMA machine or when applications do not scale up to the total number of cores.
\end{abstract}


%%%%%%%%%%%%%%%%%%%%%%%%%%%%
\section{Introduction}\label{sec:introduction}

%1. NUMA is emerging as the norm (edit joao)
Parallel architectures with non-uniform memory access (NUMA) are emerging as the norm in high-end servers.
%to do: a couple of sentences emphasizing why this trend is more and more relevant
In a NUMA system, CPUs and memory are organized as a set of interconnected nodes, where each node
typically comprises one or more multi-core CPUs as well as
one or more memory controllers. % 
%Each memory controller provides both local and remote threads with access to a partition of the
%physical address space.
The non-uniform memory access nature stems from this organization, since the memory access bandwidth (BW) and latency depends on the node where the accessing thread runs and on the node where the target page resides.

When one deploys a parallel application on a NUMA system, its threads
allocate and access pages that need to 
be physically mapped to the available NUMA nodes. 
This raises a crucial question: \emph{where should each page be mapped for optimal performance?}
%As previous research has shown, an unsuitable page placement can impact the performance of memory-intensive applications by up to a factor of 3 \cite{carrefour}.
When the application is memory-intensive, a common strategy %to answer the above question 
is to %
%keep thread-private pages at the same memory node as their owner thread, while 
uniformly interleave pages across the set of \textit{worker} nodes, i.e., the nodes on which the application threads run. This strategy is based on the rationale that, for a large class of memory-intensive applications, BW -- rather than access latency -- is the main bottleneck. Therefore, interleaving pages across %multiple 
nodes provides threads with a higher aggregate memory BW \cite{asymsched}. Hereafter, let us call this strategy \uniformworkers. This is the essential approach of recently proposed runtime libraries for NUMA systems (e.g., \cite{carrefour, asymsched}), as well as the recommended or default option for prominent database systems (e.g., \cite{mysqlnuma,mongodbnuma,cassandranuma}). 

This paper starts by questioning the effectiveness of the \uniformworkers strategy in contemporary NUMA systems. Two key characteristics of \uniformworkers seem to be at odds with the %very 
systems it aims to optimize. 
First, \uniformworkers places pages at symmetric ratios across worker nodes, while the BW (and latency) of contemporary NUMA architectures is typically asymmetric across nodes \cite{asymsched}. 
Second, there are  important scenarios where an application's threads are clustered together on a \emph{subset} of NUMA nodes -- notably, when the application is deployed in a given node partition of a co-scheduled system \cite{callisto}, or when the application does not scale beyond a subset of the available cores \cite{estima,nucore}. If the remaining memory nodes are idle or underused (e.g., by %local 
CPU-intensive applications), \uniformworkers will neglect an important portion of BW.
Hence, as we show in Section \ref{sec:motivation}, it is %not 
unsurprising that the %throughput that 
memory BW attained by \uniformworkers is 
considerably suboptimal for memory-intensive applications.

%\raggedbottom
To overcome these inefficiencies, %this paper proposes
we propose \name, a novel BW-aware page placement 
%tool 
for memory-intensive applications on %contemporary
NUMA systems. In contrast to \uniformworkers, \name takes the asymmetric BWs of every node into account to determine and enforce an optimized app\-li\-cation-specific \emph{weighted interleaving}. 
Our proposal is inspired by recent research for hybrid memory systems \cite{batman,gpu,BA-ics}.
These works have shown that, when a CPU (or GPU \cite{gpu}) is served by different memory technologies (such as NVRAM or DRAM) with differing BWs, an optimal placement is one that (proportionally) place fewer pages at the lower-BW memories.

Still, applying the same principle to the context of NUMA systems is far from trivial.
While previous BW-aware proposals for hybrid memory systems relied on the
premise that a given memory node provides the same BW to every core, that is no longer true in a NUMA system. The same NUMA memory node may be accessible through different BWs by different threads, depending on each thread's location within the NUMA topology. This implies that optimizing page interleaving from the perspective of a given worker node (as done by the recent proposals for hybrid systems \cite{batman,gpu,BA-ics}) will not always yield the best overall performance. Instead, the optimization problem needs to consider a complex $W\times N$ BW matrix, where $W$ and $N$ denote the number of worker nodes and total nodes, respectively. Furthermore, this BW matrix is particularly hard to determine accurately, since it is sensitive to interconnect congestion and local-remote contention on memory controllers phenomena which, in turn, depend on the memory demand patterns of the deployed application(s). Hence, optimal placements are eminently application-specific.

Putting it all together, an efficient page placement for asymmetric NUMA systems requires tuning $N$ weights, taking into account complex phenomena that depend both on the underlying NUMA architecture and the application(s) itself.
A naive approach is to search through the $N$-dimensional space of possible weight distributions and measure the performance of each run to find the optimal placement. This is often impractical for NUMA systems of 4 nodes and beyond, since it easily falls in the range of hours or days to find an optimized distribution of per-node weights for a given application.
An alternative approach %would be 
is to model the %BW usage 
usage of the memory system BW and analytically determine the optimal page placement.
However, to the best of our knowledge, the most successful analytical models of memory throughput %have not reached beyond 
are limited to single-node scenarios \cite{dramon}.

\name tames this complexity by combining techniques from the two extremes of the solution space. %\name relies on a novel best-effort approach that is able to find close-to-optimal per-application page placements with low overheads.
In a first stage, \name builds %an approximate 
a memory BW model of the target system. From this model, \name calculates the optimal weight distribution that maximizes the performance of a reference \emph{BW-intensive} application.
The key insight behind \name is that, after analytically determining that \emph{canonical} weight distribution, that distribution can be adjusted to fit the target application by applying a scalar coefficient on each weight.
In other words, \name reduces what in theory is an $N$-dimensional optimization problem to the one-dimensional problem of finding an appropriate scaling coefficient that best fits the %target 
application. 
To achieve this, the second stage of \name relies on an iterative technique, which, when the application starts, places its pages according to the canonical weight distribution; then, on-the-fly, it uses an incremental page migration scheme that adjusts the weight distribution until a new (local) optimum is found.

\name is implemented as
%This paper proposes \name (asymmetry-aware page placer), 
an extension to 
Linux %'s original 
\emph{libnuma}. It enriches the original interface with a \emph{bw-interleaved} policy option
%which 
that automatically determines 
%at which set of memory nodes the application pages should be placed
memory nodes to place the application pages on, 
and the %appropriate 
per-node weights to balance the page interleaving across the NUMA nodes.
\name is readily available and can be used transparently by any application, with no changes to 
the OS kernel.

%In summary, this 
This paper makes the following three main contributions.

%\begin{enumerate}
% \textit{1.} 
1. We empirically study the performance of different page placement strategies on a range of memory-intensive applications  on commodity NUMA machines. Our findings %shed new light on the page placement problem, 
show that common practices that rely on the obsolete assumption of a symmetric architecture are largely suboptimal on contemporary NUMA systems. %, and ii) expose unexplored opportunities towards better page placement strategies.

% \textit{2.} 
2. We propose \name, an extension of the \emph{libnuma} library that relies on a novel combination of analytical modelling and on-line iterative tuning.% a better page placement for %BW-intensive 
%memory-intensive applications on NUMA systems. %\name relies on a novel technique that combines analytical modeling methods with on-line iterative tuning to reach closer to the optimal page configurations that we found in our preliminary study.

% \textit{3.} 
3. We  evaluate \name on a diverse set of memory-intensive workloads, showing that \name achieves up to $4\times$ speedup %when 
compared to the Linux default \emph{first-touch} policy. %(the default in Linux).
This represents %a 
$66\%$ improvement over the performance gains that the most commonly used placement policies attain over the same baseline. These benefits are %especially prevalent
particularly relevant in scenarios where multiple co-scheduled applications run in disjoint partitions of a large NUMA machine or when applications do not scale up to the total number of available cores.
To the best of our knowledge, this is the
first proposal for BW-aware page placement in heterogeneous memory systems %to be 
evaluated on real commodity machines, i.e. not %based on 
by simulation \cite{batman,gpu,BA-ics}.

%\end{enumerate}

\begin{figure}[t!]
\begin{subfigure}{.5\textwidth}
  \centering
  \includegraphics[width=\textwidth]{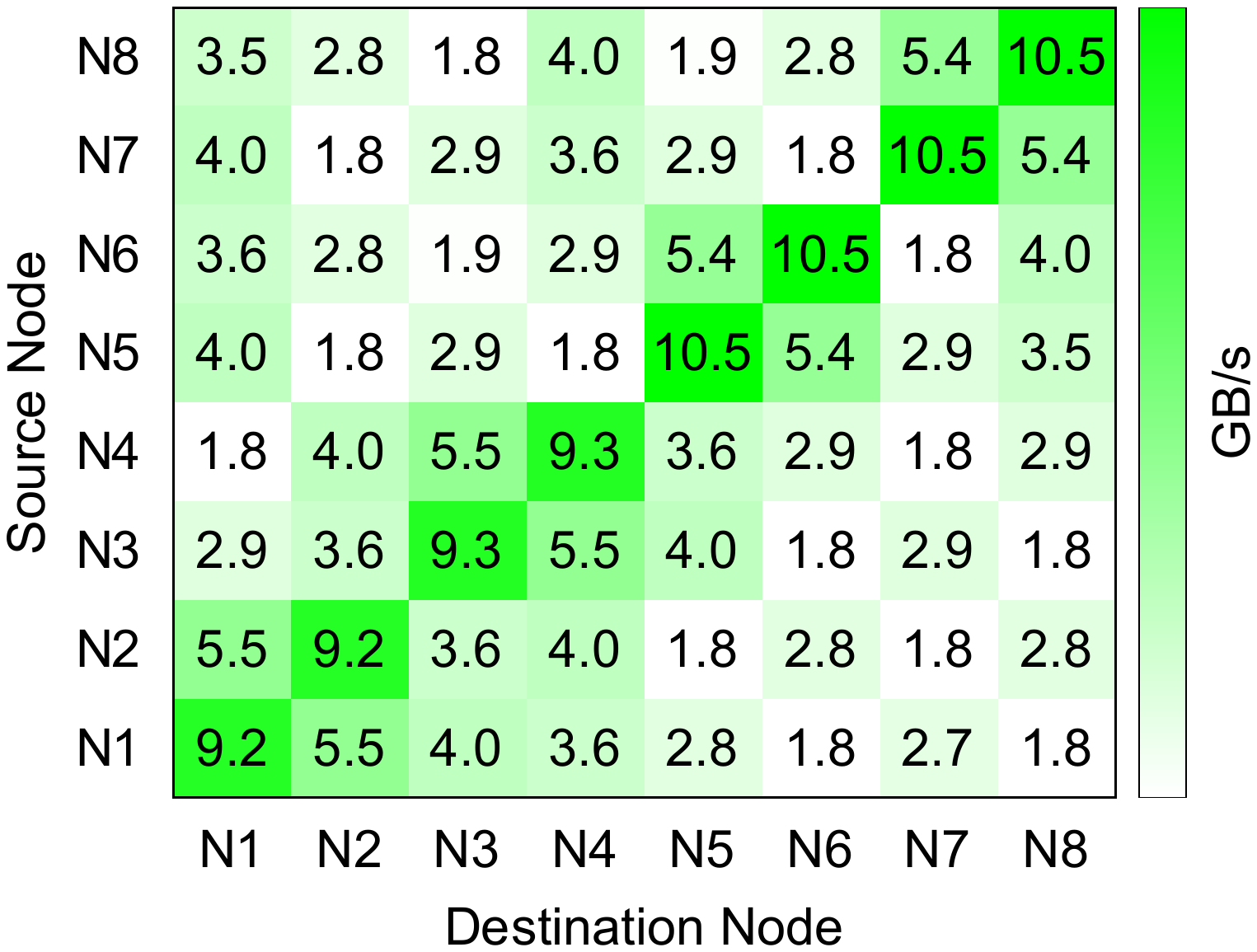}
  \caption{}
  \label{fig:bwA}
\end{subfigure}%
\begin{subfigure}{.5\textwidth}
  \centering
  \includegraphics[width=\textwidth]{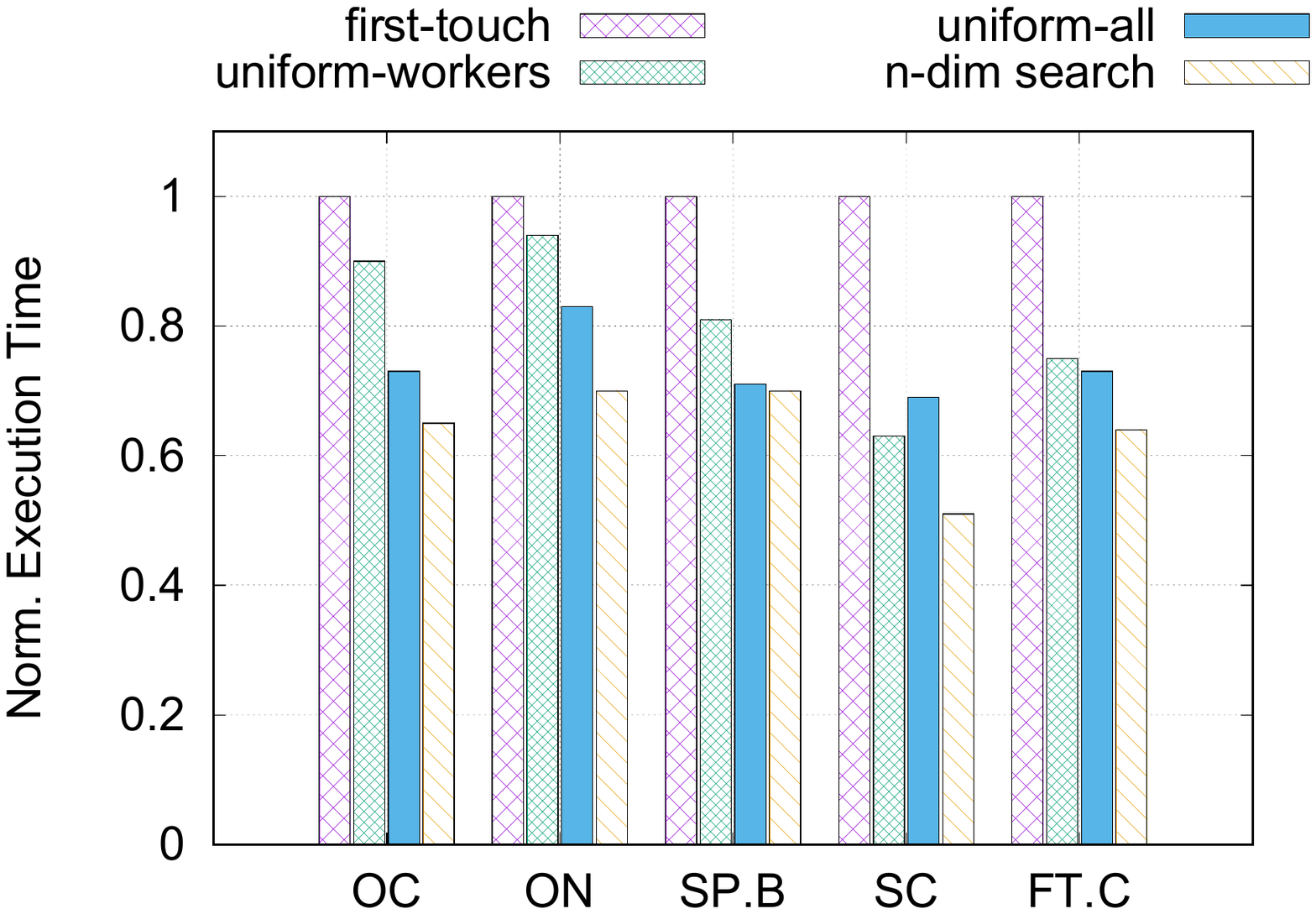}
  \caption{}
  \label{fig:n-dimexectime}
\end{subfigure}%
% \begin{subfigure}{.5\textwidth}
%   \centering
%   \includegraphics[width=\textwidth]{images/motiv-weights-2w.pdf}
%   \caption{}
%   \label{fig:n-dimweights}
% \end{subfigure}%
\caption{(a) Node-to-node BWs (GB/s) on a 8-node AMD Opteron. (b) Performance of popular page placement schemes vs the placement found via n-dimensional search for  Ocean\_cp (OC), Ocean\_ncp (ON), SP.B, Streamcluster (SC) and FT.C \cite{nas} (2 worker nodes, 8 threads each).}% on the same machine, comparing popular page placement with the best page placement found by our n-dimensional search.}% \textbf{(b):} The corresponding per-node weights as found by our n-dimensional search.}
\label{fig:n-dimbothfigures}
\squeezeup
\end{figure}

\section{Motivation}
\label{sec:motivation}

To lay the groundwork, we selected different memory-intensive applications from the PARSEC \cite{parsec}, SPLASH \cite{splash}, and NAS \cite{nas}
benchmark suites and experimentally studied how different page placement strategies affect their performance.
We used an 8-node NUMA machine with the asymmetric interconnect topology depicted in Figure \ref{fig:bwA},
on which each application ran stand-alone on 2 worker nodes.  A larger number of  architectures and baselines is evaluated in Sec.~\ref{sec:evaluation}.

For each application, we measure its  performance when its pages are mapped with Linux default \emph{first-touch}, the common practice \uniformworkers, and a \uniformall variant that uniformly interleaves pages across all nodes (both workers and non-workers). 
We also performed a long offline %iterative 
search in which we experimentally tested the performance of a large sample of weight distributions. The search used the hill climbing technique to explore the 8-dimensi\-onal space of possible solutions. 
Each point in the search space assigns to each memory node
in the machine a weight that determines the portion of pages that will be placed at that node.
The starting point in the search was \uniformworkers. %, which translates to assigning even weights
%to each worker node, and null weights to the remaining nodes. 
Each search covered approximately 180 iterations, taking more than 15 hours to complete for each application. For each application, the search identified a number of slightly different configurations that achieved performance within less than 3\% from optimum. %converged to plateaus with many diverse local optimum points that achieved similar performance (less than 1\% difference). 
Thus, the values discussed next are averages %\footnote{The resulting "averaged" configurations were confirmed to remain within 3\% from optimum.} 
over a selection of the top-10 best performing distributions for each application.
% The search eventually converged to a local optimum point (on average, after approximately 160 iterations, taking more than 15 hours to complete). 

Figure \ref{fig:n-dimexectime} presents the performance  of the baseline policies normalized with respect to that of hill climbing.
%performance of how the best performance obtained by hill climbing relates to the baseline policies. 
The results suggest that, while \uniformworkers and \uniformall considerably improve performance over Linux's default policy, they do not take full advantage of the BW of the underlying memory architecture in our NUMA system. 
To understand why, we have studied the actual weight distributions obtained by hill climbing and drawn the following three main observations, which  guided us towards the design of \name. %We next summarize these observations.

\textbf{Observation 1: Pages are placed across all nod\-es, not just worker nodes.}
In many modern architectures, even applications that have moderate single-thread memory demands can easily saturate the local memory controller when multiple threads share the same node. %multiply such demand.
%Furthermore, as Section \ref{sec:evaluation} shows, 
This issue is further exacerbated if the application threads span across multiple NUMA nodes, 
since a fraction of accesses to pages will now be remote, thus
limited by the BW of the interconnect.
%Hence, and as 
These results suggest that, for applications with high memory demands, page placement should not be restricted to the
worker nodes; instead, the available (even if limited) BW of non-worker 
nodes should be harnessed by placing on these nodes a carefully selected fraction of the application's pages.

\textbf{Observation 2: Pages are interleaved unevenly across nodes, with relevant cross-application variations.}
Every weight distribution obtained by hill climbing is highly asymmetric, with nodes with lower memory access throughput receiving fewer pages.
This clearly reflects the inherent asymmetry of the underlying NUMA topology \cite{asymsched}.  
However, when we compare the best weight distributions found for different applications, we observe significant differences between applications, which we can explain by two main factors.
On the one hand, the complex contention effects, both at the interconnect and memory controller \cite{dino}, %that impact the effective memory performance in a NUMA system 
depend on the actual memory demand that the application places on each memory node. On the other hand, while memory BW is the  dominant bottleneck of some applications, others are more sensitive to memory latency. The former benefit from exploiting the BW of remote nodes to its full extent; the latter call for approaches that, while spreading some pages remotely for increased BW, retain most pages locally for the sake of latency.

\textbf{Observation 3: If one considers worker nodes and non-worker nodes separately, proportional similarities emerge among per-node weights.} 
Let us pick two applications from our sample and compare the respective worker weights, as obtained by hill climbing, node by node. If we multiply the weights of one application by some scalar coefficient such that its aggregate worker weight becomes the same as the other application, the per-node weight variance decreases. The same occurs by following the same procedure for the non-worker nodes.
In fact, if we cluster worker nodes and non-worker nodes separately and perform the above scaling
on their (clustered) weight distributions, the average per-node coefficient of variation decreases by 1/3.

These key observations enable us to build a practical best-effort solution to BW-aware page placement in asymmetric NUMA systems, which we describe next.

\section{\name: BW-Aware page Placement} \label{sec:solution}
%\improvement[inline]{Revise and shorten this section if possible}
Given a NUMA system %machine 
and a %target 
parallel application running on a set of \emph{worker nodes}%(possibly smaller than the full set of nodes)
, the goal of \name is to devise and enforce an efficient interleaving of the application's pages across the NUMA nodes.
Since %each 
application threads access different memory nodes through potentially diverse BWs, %the interleaving produced by \name is naturally asymmetric. More precisely, 
\name %may assign assigns distinct nodes with different \emph{weights}. 
assigns different \emph{weights} to different nodes. A node's weight denotes the fraction of pages mapped to the node.

\name pipelines %of a pipeline of 
two key components, the \emph{\bwapA} and the \emph{\bwapB}. 
The first one is agnostic of the target application and runs offline. The second one is an online component that, at run-time, departs from the first component's output to reach an improved application-specific page placement. 

The inner workings of each component can be summarized as follows. %Before describing each component in detail, we give an intuitive overview.
The \bwapA models our knowledge of the BW of the NUMA topology.
Since the effective per-node BW is sensitive to the demand that applications pose on the system,
the \bwapA %conservatively 
assumes an extremely BW-intensive application as its reference.
Its output is a set of \emph{canonical weight distributions}, which optimize the memory throughput of the reference application.
If one considers different scenarios where the application runs on different sets of worker nodes, the corresponding optimal weight distributions might also differ. Hence, the \bwapA considers different combinations of worker nodes as input and %devises 
accordingly computes the corresponding canonical weights. %distributions computed.

As discussed in Section \ref{sec:motivation}, the optimal weight distribution varies substantially across distinct workloads. Therefore, the canonical weight distributions, as produced by the \bwapA, may not be suited to other applications than the idealized BW-intensive application.
Hence, for a given worker node set, the canonical weight distribution is used as a hint about the relative weight distributions to be employed for the worker set and the non-worker sets of the target application.
%Based on such relative values, \name simply needs to determine the 
%absolute weights that will be assigned to each node. 
Then, by leveraging Observation 3 (Section \ref{sec:motivation}), the \bwapB converts the canonical weight distribution to one that is optimized for the target application. 
This is done by finding an appropriate value for a \emph{data-to-worker proximity factor} (\DWP), which determines how many pages will be
assigned to the set of worker nodes (retaining the canonical weight relations), while the remainder will be shared within the non-worker nodes (also according to the canonical distribution).
%In other words, we reduce what in theory is an $n$-dimensional optimization problem to a one-dimensional one. This is the main insight that makes \name a practical solution.
The \bwapB achieves this through an incremental page migration mechanism that searches for a good value of \DWP for the target application. This 
stage is done on-the-fly, during the first seconds of the actual execution of the application.
% , with a As we show next, it is a lightweight technique with a low impact on the performance of the application.

The following sections detail each component of \name.
\subsection{Calculating canonical weights}
%The first step of the \bwapA is to model the underlying machine and an idealized canonical application running on it. Based on this model, the \bwapA determines the canonical weight distribution.
First, the \bwapA models the underlying machine and an idealized canonical application running on it. Using the model, the \bwapA computes the canonical weights.

\subsubsection{System model}
\label{sec:sysmodel}

The \bwapA makes a set of simplifying assumptions on the underlying machine and the target applications.
We show in Section \ref{sec:evaluation} that this model suffices for \name to be an effective best-effort approximation, even when the assumptions we introduce next are not entirely met in reality.
We assume a cache-coherent NUMA machine comprising a set of $N$ nodes, $\mathcal{N}=\{n_1, n_2, ..., n_{N}\}$, where the set of nodes as a whole is entirely managed %entirely managed 
by a single instance of an operating system.
Each node contains one or more multi-core CPUs, which globally provide $C$ hardware threads. %together are able to run up to $C$ hardware threads. 
For simplicity, we assume that every node's local computing and memory resources (like CPU frequencies, number of cores, or local memory BW) are identical. 
Furthermore, each node includes one or more memory controllers providing access to the local memory of that node.
We abstract the memory controllers that
are local to a given node as one single-channel memory controller,
whose BW is the aggregate BW of each channel/memory controller in the real topology.

Threads %(running at any node)
may read and write pages that reside on the local node's memory, and on any remote node's memory.
In the latter case, the read/write request is sent through the interconnect, which provides full connectivity among all nodes.

%In contrast to the node homogeneity, 
The interconnect topology is asymmetric, i.e., a thread running on a given node will observe different BWs and latencies, depending
on the memory node it is accessing.
%when reading from or writing to the memory at %different nodes.
The most obvious difference in BW and latency is between local and remote accesses. 
%The former are served at high BW and low latency by the local memory controller. The latter ones, conversely,  reach a remote memory controller via the interconnect, incurring lower BW and higher latency.
Among remote accesses, different BWs and latencies may be observed, since distinct interconnect links may have distinct BWs (possibly distinct BWs for each communication direction), and paths between nodes can differ in number of hops (some nodes
are directly connected, while others communicate through multi-hop paths).

%Assumptions on the application
On top of the NUMA system characterized so far, 
our model assumes a simplified parallel application, which we refer to as \emph{canonical application}, that uses $t$ threads, where $t \leq C \times N$. Threads are placed
at a subset of nodes, $\mathcal{W} \subseteq \mathcal{N}$, which we call \emph{worker nodes}.
For simplicity, we assume that $t$ is a multiple of the number of worker nodes, and %, therefore, 
the threads of the canonical application 
are evenly distributed across the worker nodes.
We assume that parameters $t$ and $\mathcal{W}$ have been previously
tuned by some existing tool(s) for optimal parallelism tuning (tuning of $t$) and thread placement (tuning of $\mathcal{W}$), e.g., \cite{asymsched,nucore}.
%We also consider that
In addition, the above parameters do not change while the application runs, and no other processes are co-located in the canonical application's worker nodes.

Further, we consider that %all threads run a homogeneous workload. More precisely, 
the work performed and the memory access patterns are similar among all the threads of the application.
Within the application's address space, we distinguish between thread-private pages and
shared pages. We assume that the access volume to thread-private pages is negligible
when compared to the available memory BW.

In contrast, we assume that the canonical application places an extreme memory demand on the shared pages.
Moreover, the workload of the canonical application is BW-intensive, %in the sense 
such that the memory access throughput that it attains is the dominant 
factor to its overall performance, rather than memory access latency (and other overheads unrelated to memory). Hence, %from now on 
hereafter we focus on memory throughput to shared pages, and omit latency from our equations.
Furthermore, we assume that the canonical application accesses shared data predominantly in read-only mode; i.e., write accesses to shared pages are so rare that they have no relevant impact on performance. Finally, we consider that the canonical application accesses all shared pages with the same probability. 

Shared pages are interleaved across all memory nodes in a weighted fashion, where some nodes %will (possibly) 
may receive more pages than others.
No matter which page interleaving is chosen, we assume the shared space fits the available physical memory.
Pages are interleaved according to a weight distribution, $\mathcal{D} = \{w_1, w_2, .., w_N\}$,
where $w_i$ denotes the fraction of shared pages that node $i$ will hold (such that $\sum w_i = 1$).
Therefore, to maximize the performance of the canonical application, we need to find the optimal weight distribution that maximizes the overall throughput to shared data. %We address that goal next.
%Shared pages, however, are hard to place optimally, as they are accessed by threads residing on different nodes, interconnected by an asymmetric topology.

%paolo: Optimized weight distribution => Canonical weight distribution ?
\subsubsection{Finding the optimal weight distribution}
\label{sec:model:canonicalweights}

Now we discuss how to find the optimal weight distribution for the canonical application. %that 
%maximizes the canonical application's performance. 
We introduce the function $bw(n_{src} \rightarrow n_{dst})$, which
denotes the BW that a given thread located at worker node $n_{dst}$ can use when reading from a  node $n_{src}$. % or, conversely, that a thread at $n_{src}$ can use when writing to $n_{dst}$.
For presentation simplicity, we start by assuming that function $bw(n_{src} \rightarrow n_{dst})$ is well known for all node pairs in the system and does not change for different weight distributions. We discuss how to lift these assumptions in the next section. %We will shortly extend our solution to more realistic scenarios where we drop each of these simplifications.

\textbf{Single-worker scenario.} The simplest scenario is where all threads  run on a single worker node, $n_w$. 
%Our solution 
\name adapts the approach that Yu et al. proposed in the context of data placement in SMP systems with heterogeneous DRAM-NVM memory hierarchies \cite{BA-ics}.
We start by considering that the threads in $n_w$ need to read a total of $S$ bytes of shared data before completing their work.
Since the canonical application's performance is dominated by memory
throughput, its execution time is  determined by the time that the threads in $n_w$ take to transfer $S$ bytes from memory. Since memory is interleaved among the memory nodes of the system, threads will need to read from pages held at distinct nodes. Furthermore, since the canonical application accesses any page with a uniform probability, the portion that is read from each node $i$ will be proportional to the weight of $i$; more precisely, $S \times w_i$ bytes.

Moreover, we consider that the data sets that $n_w$ reads from each node in the system are transferred in parallel to $n_w$. %
%Since node $n_w$ runs many parallel threads, we assume that the data stream every node is read in parallel the the reads from portion bytes transferred from each node in the system to $n_w$ occur in parallel. 
As Yu et al. have shown \cite{BA-ics}, this approximation is relatively accurate for memory-intensive applications in systems where the number of accessing %hardware 
threads in a node is substantially higher than the number of source memory nodes. %In other words, given any node $n_i$, we consider that, at any instant, there is at least one thread in $n_w$ currently reading from node $n_i$.
%Our experimental results in Section \ref{sec:evaluation} confirm this observation.

Putting it all together, we can then determine the execution time of the threads in $n_w$ as the time to complete the longest (parallel) transfer from every memory node:

\vspace{-1em}
\begin{equation}
T = \max_{i\in \mathcal{N}}\frac{S \times w_i}{bw(n_{i} \rightarrow n_{w})}
\end{equation}

If pages were uniformly interleaved (i.e., equal weights for all pages), execution time would be determined by the time to transfer from the node with the lowest BW (to $n_w$). Hence, to minimize execution time, one can reduce the pages placed in that node (by decreasing the corresponding weight) until it no longer incurs the longest transfer time. After that, another node becomes the one that contributes with the longest transfer time and its weight should also be reduced; and so forth. It is easy to show that the resulting optimal solution consists of setting the weight of each node $n_i$ as follows: 

\vspace{-1em}
\begin{equation} 
w_i = \frac{bw(n_i \rightarrow n_w)}
{\sum_{i\in \mathcal{N}} bw(n_{i} \rightarrow n_{w})}
\end{equation}

\textbf{Multi-worker scenario.} 
We now generalize the previous solution to scenarios where the canonical application has a higher parallelism level, thus spans its threads across two or more worker nodes.
In this scenario, the execution time of the application is given by the time that (the threads running at) the slowest worker node takes to complete, as follows:

\vspace{-1em}
\begin{equation} 
\label{eq:multiworker1}
T = \max_{n_w \in \mathcal{W}} ( \max_{i\in \mathcal{N}}\frac{S \times w_i}{bw(n_{i} \rightarrow n_{w})})
\end{equation}

Like in the single-worker scenario, we can minimize the execution time by adjusting the weight distribution. 
However, the multi-worker scenario introduces a subtle challenge: %since 
the BW that two distinct worker nodes, $n_A$ and $n_B$, may use when reading from a target node $n_i$ can be different. In fact, changing the weight of $n_i$ to greedily optimize $n_A$'s performance can degrade $n_B$'s performance, and vice-versa.
Therefore, the optimization of the weight distribution needs to take all the transfers by all the worker nodes into account.

We achieve this by defining %the notion of 
\emph{minimum BW} as the BW of the weakest path among the paths interconnecting a given target 
node to the %set of 
worker nodes as $minbw(n) = \min_{n_w \in \mathcal{W}} bw(n_{i} \rightarrow n_{w})$, and by transforming Equation \ref{eq:multiworker1} to employ this notion:

\vspace{-1em}
\begin{equation} 
\label{eq:multiworker2}
T = \max_{i\in \mathcal{N}}\frac{S \times w_i}
{minbw(n_{i})}
\end{equation}

Finally, we can apply a similar optimization strategy as we do in the single-worker scenario, but this time by considering the minimum BW values. Consequently, %it is easy to show that 
the optimal solution in the multi-worker case consists of making each node's weight proportional to its minimum BW:%(relatively to the worker node set):

\vspace{-1em}
\begin{equation} 
w_i = \frac{minbw(n_i)}
{\sum_{i\in \mathcal{N}} minbw(n_{i})}
\end{equation}

% The following sections describe how we can employ the above solutions when the throughput function is not known \emph{a priori} and when other accesses than reads to shared pages have a relevant impact on performance.

\subsubsection{Estimating BW}
\label{sec:model:bw}

The solution built so far assumes that the BW between two nodes ($bw(n_{src} \rightarrow n_{dst})$) is well known and does not change for different weight distributions. We now question these assumptions and propose a practical solution that does not depend on them.

It is important to understand what main factors contribute to the BW that threads in a NUMA machine may effectively use.
%Of course, 
The value of $bw(n_{src} \rightarrow n_{dst})$ is limited by the nominal BW of the %interconnect 
path from $n_{src}$ to $n_{dst}$ (if $n_{src}\neq n_{dst}$). %if both nodes are distinct) and by the BW of the local memory controller at $n_{src}$. 
However, the effective BW that threads in $n_{dst}$ get when reading from $n_{src}$ is also determined by the access demand that the application places on the system, in different complex ways. 

First, it is %well 
known that
the actual BW of a single memory controller is influenced non-linearly by the actual access demand to that %memory
controller \cite{dramon}. To complicate matters, the effective BWs, as perceived from %different 
worker nodes, are also strong\-ly affected by cross-thread and cross-node interference \cite{denoyelle:hal-01622582}.
%On the one hand, 
The access demand on the memory controller at node $n_{src}$ includes concurrent accesses %issued by 
from multiple threads running in the same worker (either $n_{src}$ or a remote node); further, if %or, in case
the canonical application spans multiple worker nodes, the memory demand on $n_{src}$ combines contending accesses by threads residing on distinct nodes (besides local threads at $n_{src}$). 
Finally, BW is also affected by interconnect congestion. This may happen when multiple co-located threads access the same remote memory node (through the same link), or when threads from different worker nodes issue memory requests whose result is delivered through interconnect paths that share one or more links. %
%joao: I removed these complementary explanations, which were quite nice. Consider reusing them if we find free space.
Therefore, it is clear that our initial assumptions on $bw(n_{src} \rightarrow n_{dst})$ are questionable. First, we relied on an exact knowledge of BW in a NUMA machine. %However, to the best of our knowledge, the most accurate memory BW models proposed in literature have have not reached beyond single-node scenarios \cite{dramon}. 
Second, when searching for the optimal weight distribution, we assumed that memory throughput was immutable during that search.

\name follows a pragmatic approach to approximate 
the $bw(n_{src} \rightarrow n_{dst})$ function. %by iteratively profiling the real memory throughput of the machine over a few weight distributions. 
More precisely, for a fixed set of worker nodes, we start by deploying a memory-intensive benchmark (representing the canonical application) and uniformly interleaving the benchmark's pages across all nodes in the machine.
We use a simple benchmark that spawns
as many threads as the available hardware threads on the worker nodes, and each thread performs a random traversal of a shared array. At the same time, we rely on hardware performance counters to monitor per-node memory throughput. 
%joao: maybe provide a bit more details about the perf monitoring 

The profiled throughputs between each pair of nodes, $n_{src}$ and $n_{dst}$, are used as the values of $bw(n_{src} \rightarrow n_{dst})$. % when \name selects the weight distributions by employing the equations in the previous section. 
This approach %is inaccurate as it 
neglects the differences in access demand that occur when page placement changes from the profile-time uniform interleaving scenario to the final weighted interleaving scenario. However, our %experimental 
results (in Section \ref{sec:evaluation}) confirm that, %despite this inaccuracy, 
\name is still able to devise efficient weight distributions.

On the practical side, at installation time on a given machine, the \bwapA  needs to run the above profiling procedure for the relevant combinations of worker node sets (with different sizes). The set of explored worker node sets  does not need to be exhaustive: i) a large number of worker node sets can be filtered out since they are unlikely to be used by a rational user (e.g., in a dual-socket machine with 2+2 nodes, a 2-worker set comprising nodes at each socket, thus interconnected with a low BW); ii) many worker node sets are symmetrical, hence only one needs to be configured (e.g., in a dual-socket machine with 2+2 nodes with symmetric links between sockets, the optimal weight distribution for the worker set comprising two nodes on one socket is symmetrical to the set comprising the nodes on the other socket).

\subsection{On-line page placement tuning}\label{sec:adaptive}
%\info[inline]{Add a brief section about the coordinator for the co-scheduled scenario}

%intro
The \bwapB takes action when an application is launched.
%This component consists of a library with 
The tuner's API includes a main function \texttt{\name-init}%. This function 
, which should be called by the target application once it has allocated its initial shared structures. 
%\hl{
%We 
Note that \mbox{\bwapB} targets applications, %that
which, after an initial stage, enter an execution stage with stable memory
access behavior (identically to systems such as Carrefour \cite{carrefour} and Asymsched \cite{asymsched}). %}
The main argument of \texttt{\name-init} %indicates
points to the set of worker nodes on which the  application is running.

The goal of the \bwapB is to tune the  weight distribution by searching for an appropriate application-specific \DWP. We recall that \DWP determines the balance between pages mapped to the set of worker and non-worker nodes, while preserving the relative weight relations within the sets of worker and non worker nodes. %among worker nodes, as well as among non-worker nodes. 
The canonical weight distribution corresponds to  $DWP=0$, while $DWP=1$ corresponds to the extreme where all pages are mapped to the worker node set.
This  allows \name to be used with both BW-sensitive (low \DWP) and latency-sensitive workloads (high \DWP). 

\subsubsection{Tuning data-to-worker proximity}\label{subsec:library}

%The \bwapB operates as follows.
Initially, the \bwapB obtains a pre-computed canonical weight distribution for the worker node set. The %already 
application's shared pages  are initially placed based on this weight distribution ($DWP=0$).

The \name library adapts \DWP on-the-fly during the first seconds after the application calls \texttt{\name-init}, based on hill climbing. Departing from the $DWP=0$, we periodically monitor average resource stall rates (stalled cycles per second), by reading  hardware performance counters via a portable library \cite{likwid}. It is well known that stalled cycles are strongly correlated to execution time \cite{estima}. 
At each period, we collect $n$ measurements over an interval of $t$ seconds. We then sort and discard the first and the last $c$ measurements  to filter outliers. At each iteration, we compare the current average stall rate with the previous one, and accordingly vary \DWP by a constant step, $x$. %In our current implementation, we set $n$ to 20, $c$ to 5, $t$ to 0.2, and $x$ to 10\%.
If the stall rate decreased, %relatively to the previous period, 
it is likely that  increasing \DWP improved the overall performance. Hence, we continue increasing \DWP. Otherwise, %at the end of some period, stall rate increased or stagnated relatively to the previous period, then 
it is likely that we found a (local) optimum and we stop the hill-climbing. % and keep the current weight distribution.

At each iteration where we decide to increase \DWP, applying the new value is ensured by incrementally migrating pages from non-worker nodes to worker nodes, where the number of pages that is removed from/added to a given node is proportional to that node's canonical weight. This ensures that the relative canonical weights within the worker and non-worker node set %and the non-worker set 
are preserved as \DWP changes, as intended.

\subsubsection{Initial placement and incremental migration}
\label{sec:pageplacement}

At each iteration, the \bwapB needs to place pages according to a weighted interleaving strategy.
However, no direct support for \emph{weighted} interleaving exists in mainstream OSs.
Hence, we have to build such support for weighted-interleaved page placement. We achieve this by two alternative means: at kernel and user levels.
%As the first alternative, 
At the kernel level, we implemented a new policy to support weighted interleaved memory allocation, exposed by a new system call. We also added the weighted interleave option to \emph{numactl} tool and \emph{libnuma} library to avoid the burden of application-level changes.

Our user-level alternative has the advantage of portability (avoiding the need for patching the underlying kernel), yet at the cost of a less accurate interleaving. Algorithm \ref{alg:algo} summarizes our user-level page placement algorithm. 
We start by determining the currently allocated page ranges that are likely to hold shared data. This includes the \emph{.data} and BSS segments, as well as dynamic memory mappings. We divide each address range into contiguous sub-ranges and, for each sub-range, we call \emph{mbind} with the uniform interleaving option over a different set of nodes (line 8 in Alg. \ref{alg:algo}). More precisely, the first sub-range's pages are interleaved over all nodes, the second sub-range's pages are interleaved over all nodes except the one with the lowest weight, and so forth. The key insight is that, by %carefully
setting the size of each sub-range (line 7 in Alg. \ref{alg:algo}), we can ensure that the overall per-node page ratios will be proportional to the desired weights.

This solution is not as accurate as the kernel-level alternative, since it does not enforce that all the sub-ranges reflect the weighted interleaving. Still, it ensures a best-effort shuffling of pages, %(it avoids placing large segments of consecutive pages on the same node)
while keeping the number of \emph{mbind} calls low. %As we show in \ref{sec:evaluation}, the different accuracy of each alternative do not have a relevant impact on the overall performance of the applications we studied.
By using the \emph{M\-POL\_MF\_MOVE} and \emph{MPOL\_MF\_\-STRICT} flags of \emph{m\-bind}, this approach also works correctly (without kernel modifications) when weight distributions are changed dynamically by the \bwapB. 
It is easy to show that, as \DWP increases in each step, Algorithm \ref{alg:algo} will call \emph{mbind} on sub-ranges that were previously mapped according to the same or a wider %(covering more nodes) 
interleaving. In this case, \emph{mbind} seamlessly performs the necessary page migrations. The reverse migration/operation is currently unsupported by \emph{mbind}.

\begin{algorithm}[t]
\scriptsize
%\tiny
\SetAlgoLined
\SetKwInOut{Input}{Input}
%\Input{$segment$ \tcp*[h]{Structure containing segment starting address and its length}}
%\Input{$nodes$ \tcp*[h]{Set of NUMA nodes and their canonical weight distributions}}
\Input{$segment$ \tcp*[h]{%Structure 
Struct with %starting 
start address and length}}
\Input{$nodes$ \tcp*[h]{Set of NUMA nodes and respective weights}}
\Begin{
    % the address where we start the interleaving
    $address \leftarrow segment.startAddress()$\;
    % weight of the previously selected node
    $weight_{prev} \leftarrow 0$\;
    % we go through all of the nodes in ascending order of weights
    % we interleave disjoint parts of the segment in different sets of nodes
    \While{$nodes \neq \emptyset$}{
        $node \leftarrow getNodeWithMinWeight(nodes)$\;
        % we compute the size to be passed onto mbind
        $weight \leftarrow node.weight - weight_{prev}$\;
        $size \leftarrow |nodes| * weight * segment.length()$\;
        % we call MPOL_INTERLEAVE on the computed address, size and set of nodes
        $interleave_{uniform}(address, size, nodes)$\;
        % we then remove this node from the set, update address and take note of the weight of this node
        $nodes \leftarrow nodes - \{node\}$\;
        $address \leftarrow address + size$\;
        $weight_{prev} \leftarrow node.weight$\;
    }
}
\caption{User-level weighted interleaving approximation used by the \bwapB}
\label{alg:algo}
\end{algorithm}

\subsubsection{Co-scheduled variant}\label{subsubsec:multiple-workloads}

The mechanism described so far assumed a stand-alone application that runs on a subset of nodes and also has the remaining nodes idle to place its pages.
%\hl{
However, many NUMA systems will consolidate workloads in a single physical machine in order to minimize idle hardware resources.
Workload consolidation is today an active research topic, both in academia and industry (e.g., with Intel's recent introduction of Resource Director Technology \mbox{\cite{intelrdt}} into its high-end processors). 
A common formulation of the problem considers one or more \emph{high-priority} workloads that are supposed to perform as well as if they were accessing isolated resources (e.g. memory); and one or more \emph{best-effort} workloads that benefit if provided with some resources that are originally assigned to the former workloads \mbox{\cite{parties,heracles}}.
Recent proposals to this problem \mbox{\cite{copart,emba}} focus only on single-socket scenarios. In \mbox{\name}, we improve on this by enabling the allocation of the BW of the full set of NUMA nodes across two or more applications running in disjoint worker nodes.%}

In the simplest co-scheduling scenario, 
we can consider one high-priority workload, $A$, that has a low memory intensity; and a best-effort workload, $B$, that is memory-intensive. 
It is desirable that $B$ places some of its pages on the nodes where $A$ runs (i.e., $B$'s non-worker nodes), so  $B$ can benefit from the spare BW of those nodes.
However, such memory consolidation strategy should not degrade $A$'s memory performance.
To support this scenario, the \bwapB supports a co-scheduling variant, %. In this variant, 
where the iterative search comprises two stages, both coordinated by an external process that monitors the stall rates of both applications.
The rationale of this 2-stage approach is that, below a given \DWP of $B$, the demand that A will pose on the local nodes of $A$ will start having a noticeable degradation of $A$'s performance. Hence, the search should only consider \DWP values that are above that lower bound. 

The first stage determines an approximation of such %upper 
bound. %To achieve this, 
The monitoring process measures the stall rate of $A$, increasing the \DWP of $B$ as long as the stall rate of $A$ keeps decreasing. %As soon as 
When $A$'s stall rate stabilizes, the search has found a local maximum of $A$'s performance, which is probably a good approximation of the lower bound on the \DWP of $B$.
At that point on, the second stage starts %, which 
and proceeds as described %for the stand-alone
in Section \ref{sec:adaptive}-2 (i.e., now guided by the stall rate of $B$).%}

%\hl{
As a final remark, we note that there are two aspects that the \mbox{\bwapB} does not currently handle automatically. First, in the co-scheduled variant, we assume that some external tool/hint\mbox{\cite{carrefour}} has classified each workload as memory-intensive or not. 
Second, the programmer is expected to call \mbox{\name}-init when the program is about to enter its stable phase.
The first limitation can be addressed by using the number of memory accesses per instruction (MAPI) to classify workloads as either memory-intensive or not (%similarly to what 
like in Carrefour \cite{carrefour} %does
).
As for the second limitation, one may consider looking at the periodic variation of the MAPI metric and only trigger the \mbox{\bwapB} when such variation is below a given threshold.

\section{Evaluation}
\label{sec:evaluation}

Our evaluation answers two key 
questions:
\emph{1.~What performance advantage does \name bring to %real
memory-intensive applications on NUMA systems compared to %other 
state-of-the-art page placement policies like Carrefour \cite{carrefour} or Asymsched \cite{asymsched}, which rely on uniform interleave to place shared pages?}
%architectures, when compared to other page placement policies?}
\emph{2.~Considering each main component of \name %components 
separately, %, namely, (1) the \bwapA %the canonical weight distribution and 
%(2) the \bwapB %on-line tuning of page placement
how effective is it and what overheads does it introduce?}

We consider several %of a set 
state-of-the-art page placement policies, i.e.,
%Namely, 
%Specifically, we compare the performance of
the Linux's default policy (\emph{first-touch}), uniform interleaving across workers (\uniformworkers), uniform interleaving across all nodes (\uniformall), and \emph{autonuma} \cite{autonuma}.

Among the different policies evaluated, \uniformworkers is especially relevant since it is the core strategy that state-of-the-art proposals adopt when placing pages across memory nodes of a NUMA system. %across a NUMA topology. 
Among others, this includes proposals like Carrefour \cite{carrefour}, Asymsched \cite{asymsched}, and the BW-aware policies proposed by Baek et al.\cite{BA-ics}.
Since other recent BW-aware page placement proposals do not support NUMA (e.g.
\cite{batman,gpu}) they are not represented here.%in this evaluation.

Although Carrefour and Asymsched resort to \uniformworkers, they complement it with two main optimizations: namely the detection and co-location of private pages, and the replication of read-only pages. We do not evaluate these features since they require kernel modifications and no patch was available for our operating system versions. Still, we note that such optimizations are orthogonal to our paper, since they can directly complement the mechanisms proposed in \name. 

\begin{table}[]
\resizebox{\textwidth}{!}{
\caption{Memory access characterization of the evaluated benchmarks, as obtained by NumaMMA \cite{numamma} tool on machine B running each benchmark on a full worker node.
The trends of each benchmark do not change significantly for higher number of workers (results
omitted for space limitations).}
\label{tab:t2}
\begin{tabular}{|r|r|r|r|r|}
\hline
\multirow{2}{*}{Benchmark} & \multicolumn{2}{r|}{BW Requirements} & \multicolumn{2}{r|}{Memory Access Pattern} \\ \cline{2-5} 
 & \begin{tabular}[c]{@{}r@{}}Reads \\ (MB/s)\end{tabular} & \begin{tabular}[c]{@{}r@{}}Writes \\ (MB/s)\end{tabular} & \begin{tabular}[c]{@{}r@{}}Private \\ Accesses (\%\end{tabular} & \begin{tabular}[c]{@{}r@{}}Shared \\ Accesses (\%)\end{tabular} \\ \hline
Ocean\_cp (OC) \cite{splash} & 17576 & 6492 & 79.3\% & 20.7\% \\ \hline
Ocean\_ncp (ON) \cite{splash} & 16053 & 5578 & 86.7\% & 13.3\% \\ \hline
SP.B \cite{nas} & 11962 & 5352 & 19.9\% & 80.1\% \\ \hline
Streamcluster (SC) \cite{parsec} & 10055 & 70 & 0.2\% & 99.8\% \\ \hline
FT.C \cite{nas}& 5585 & 4715 & 95.0\% & 5.0\% \\ \hline
\end{tabular}}
\squeezeup
\end{table}

Besides these alternatives, we evaluate %a full implementation of 
complete \name and an incomplete variant, denoted \emph{\name-uniform}, which disables the \bwapA. This variant departs from \uniformall as its initial weight distribution and only runs the \bwapB.

For space limitations, the weighted page interleaving in \name is enforced with the portable, user-level option. %, which is preferable over the kernel-level alternative for its portability. 
By enabling the kernel-level variant, we observed only marginal gains (at most 3\%) and reach the same main conclusions.
The parameters of the \bwapB of \name (Section \ref{subsec:library}) are set as follows: $n=20$, $c=5$, $t=0.2$, and $x=10\%$. We chose these values by optimizing the parameters for a particular setting (benchmarks Ocean\_* on machine A), then used those values for every other benchmark/machine/experiment.

%applications
For our evaluation, we have %evaluated 
selected memory-intensive benchmarks from %several benchmark suites, %i.e.namely, 
NAS\cite{nas}, PARSEC\cite{parsec} and SPLASH\cite{splash} %benchmark 
suites.
We intentionally omit benchmarks with low memory intensity, since page placement has marginal impact on their performance.
%\hl{
%We recall that, 
Although \name's design assumes read-only and shared-only pages, we %intend to 
evaluate how it performs as a \emph{best-effort} approach with workloads that do not fit those assumptions.
Therefore, our selection of benchmarks promotes diversity with regard to the ratios of read vs. write accesses, and  thread-private vs. shared accesses.
%Table \mbox{\ref{tab:t2}} details the memory access characteristics of each benchmark, 
The memory access characteristics of each benchmark is shown in Table \mbox{\ref{tab:t2}}, 
which confirms that most benchmarks are far from the assumptions underlying \name's design. 
%More 
Specifically, three of the benchmarks: OC, ON and FT.C -- have more than 79\% of thread-private memory accesses.
%memory accesses that are thread-private. %}
%Another aspect where 
Benchmarks also differ in their scalability, as Table \mbox{\ref{tbl:scenarios}} summarizes.
As for the datasets, we used the largest available datasets, i.e., the native inputs for %the 
PARSEC and SPLASH %benchmarks 
and the CLASS B and C datasets for NAS. 
%All 
Each of these datasets fits in the memory of each node. We use  execution time, averaged over 5 runs, as the performance metric. 

All the  benchmarks are multithreaded applications with a malleable thread pool.
For thread placement, we used the usual rule of thumb that is adopted, for instance, by AsymSched \cite{asymsched}: threads are grouped together %(each thread pinned to a distinct core) 
on the subset of worker nodes with the highest aggregate inter-worker BW. To minimize scheduling-related overheads (e.g., core over-subscription, simultaneous multithreading, thread migration), we pin each thread to a distinct core. %, they remain there throughout the application's execution time.
We leave the interaction of page placement and OS-guided thread scheduling as future work. %\hl{
The page size used in our experiments %and \name 
is the Linux default page size ($4$KB). %Since handling large pages %in NUMA is an orthogonal problem\mbox{\cite{large-pages}}, 
%is orthogonal to NUMA handling \cite{large-pages}, we disabled large pages. 
Integrating \mbox{\name} with large pages in NUMA systems \cite{large-pages} is left as future work.

We ran our experiments on $2$ NUMA systems of different scales and asymmetry. %Table \ref{table:t1} summarizes the detailed system specification of these systems. %which we detail next.
Machine A is a 4-socket AMD Opteron Processor 6272%@2.1GHz
, with 8 memory nodes, 8 cores per node, 64GB DRAM, running Linux 4.17.
Machine A is representative of a high-end NUMA system with a strongly asymmetric interconnect topology.
As it is evident in Figure \ref{fig:bwA}, interconnect links exhibit ample disparities in link BW, sometimes affecting even different directions of the same link. %sometimes at each direction between the same pair of nodes.
Machine B is a 2-socket Intel Xeon CPU E5-2660 v4. %@2.00GHz
In contrast, Machine B is a smaller-scale machine with a simpler topology. It has 4 NUMA nodes (using Cluster-on-Die mode), 7 cores per node, 32GB DRAM, running Linux 4.4.
%which Figure \ref{fig:bwB} presents. 
In this case, memory BW still exhibits asymmetries, however less pronounced than in machine A. 
While the lowest BW in machine A was $5.8\times$ %times 
lower than the highest BW (i.e., local memory BW), that amplitude drops to $2.3\times$ %times 
in machine B.

\begin{figure}[t!]
\setlength{\textfloatsep}{1\baselineskip plus 0.2\baselineskip minus 0.5\baselineskip}
%\hspace*{-0,3cm}
\begin{subfigure}{.375\textwidth}
  \centering
  \includegraphics[width=\linewidth]{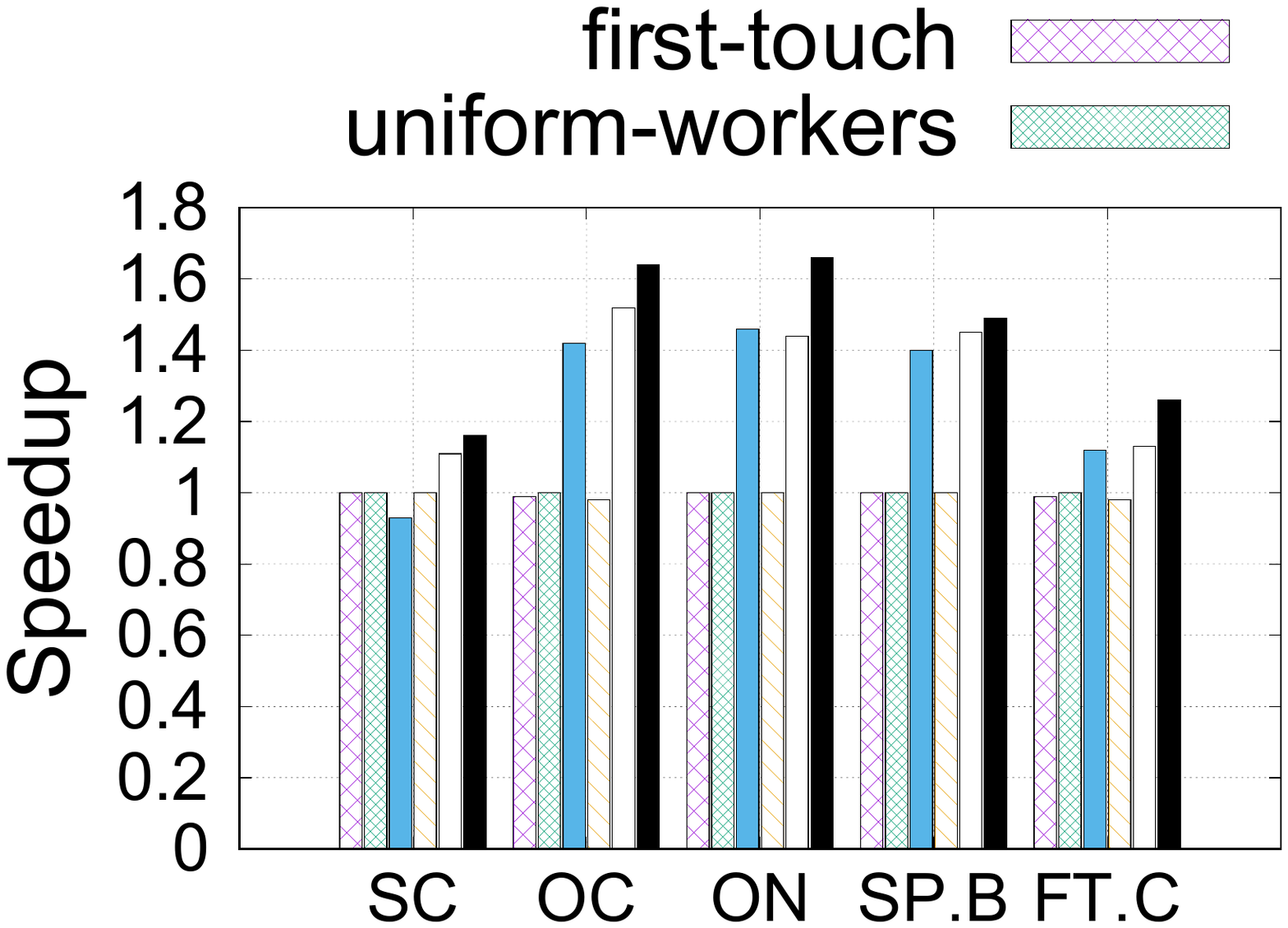}
  \caption{One worker node}
  \label{fig:machineAw1}
\end{subfigure}%
\hspace*{-0,5cm}
\begin{subfigure}{.375\textwidth}
  \centering
  \includegraphics[width=\textwidth]{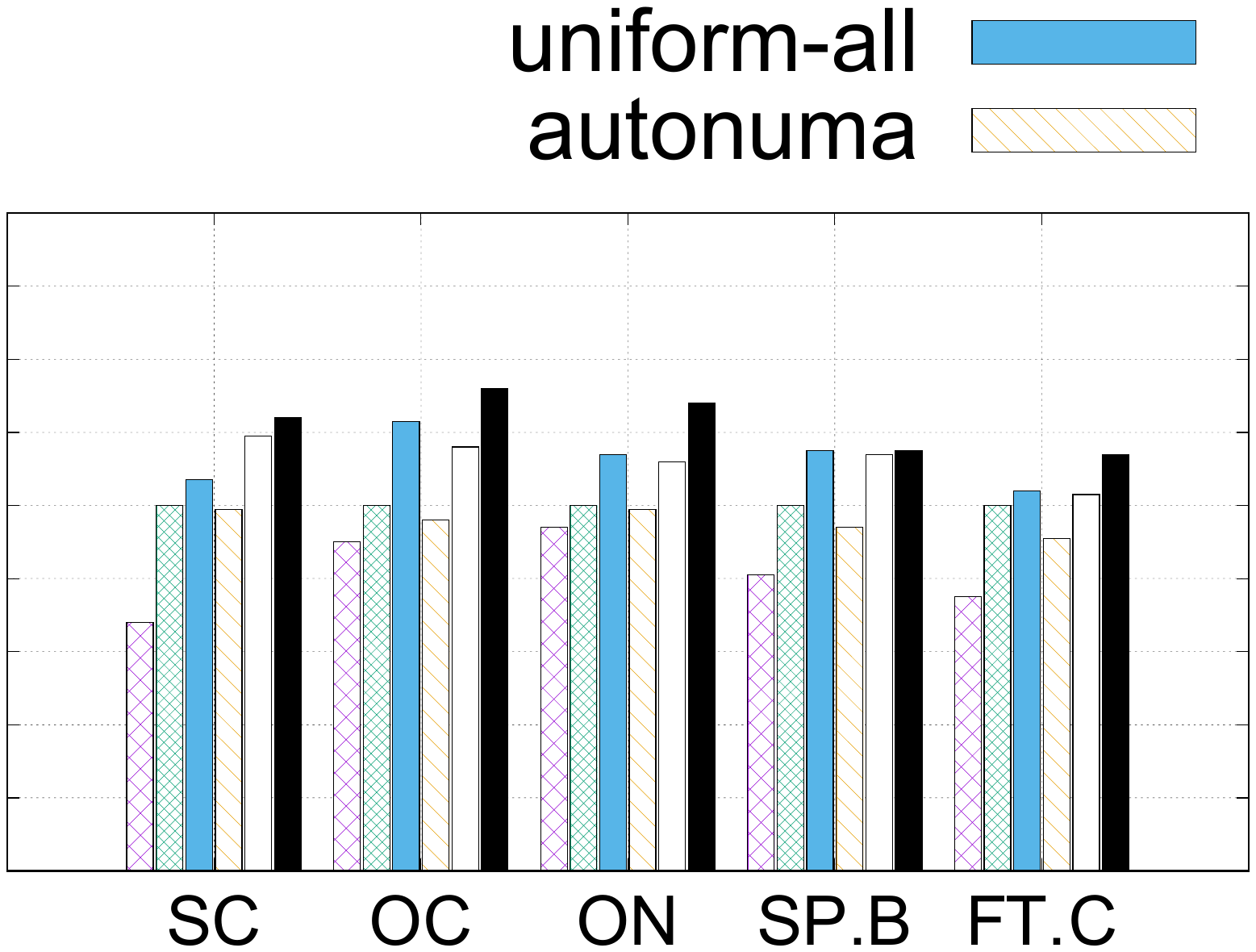}
  \caption{2 worker nodes}
  \label{fig:machineAw2}
\end{subfigure}%
\hspace*{-0,4cm}\begin{subfigure}{.375\textwidth}
  \centering
  \includegraphics[width=\textwidth]{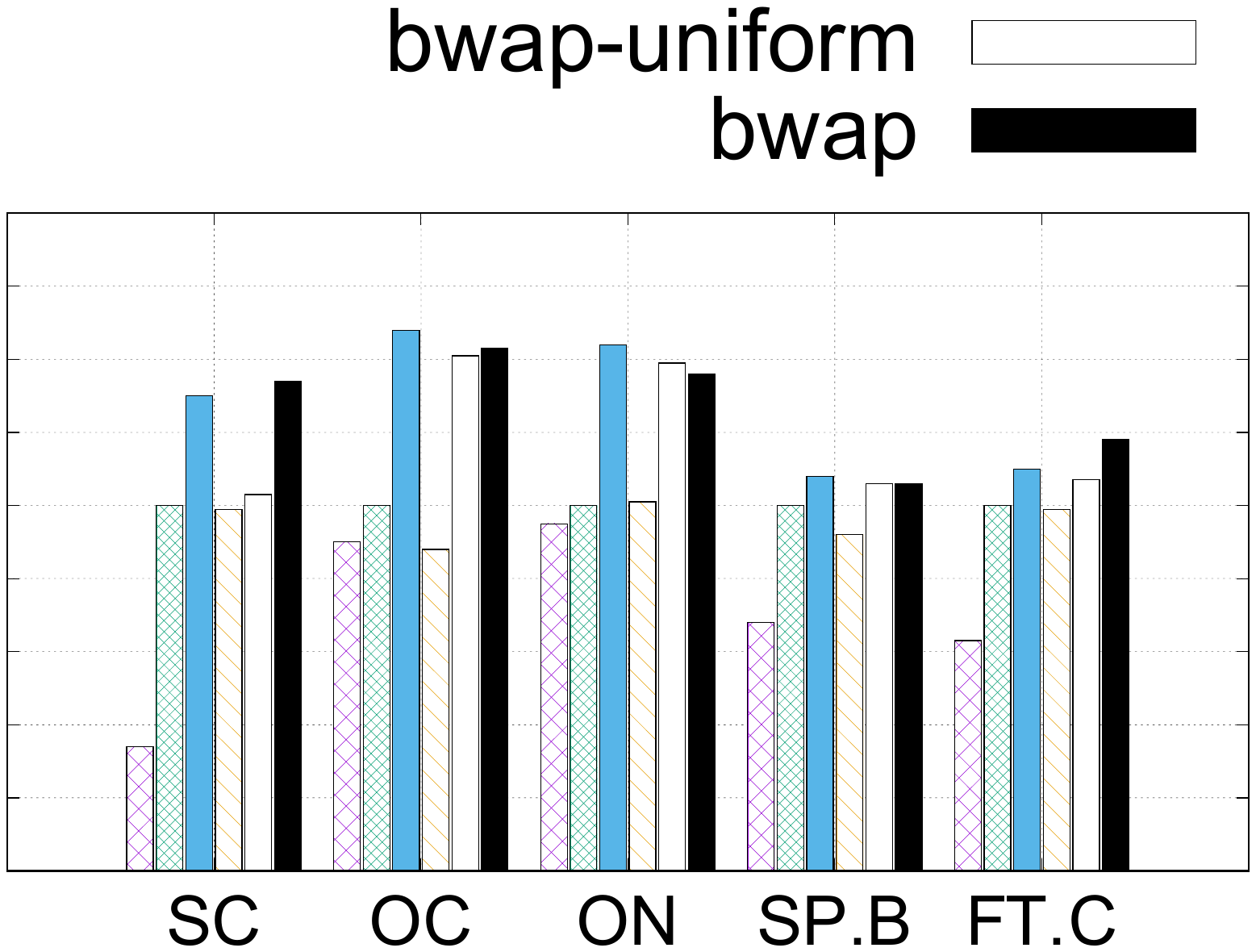}
  \caption{4 worker nodes}
  \label{fig:machineAw4}
\end{subfigure}%
\caption{Speedup vs \emph{uniform workers} (co-scheduling, mach.A)}% scenario %, on up to 1 \textbf{(a)}, 2 \textbf{(b)}, and 4 \textbf{(c)} worker node(s) 
%on machine A.}%, for different page placement solutions.}
\label{fig:detailedmachineA}
\squeezeup
\end{figure}

\subsection{Overall performance evaluation}
\label{sec:evaluation:apps}

To %quantify 
compare the performance of \name to page placement alternatives,  
%relative to the page placement alternatives, 
we measure the performance of %different %memory-intensive 
bench\-marks when using different placement policies.
%Our %initial focus goes to 
We focus on two representative execution scenarios: 
\emph{co-scheduled} and \emph{stand-alone}. 

\textbf{Co-scheduled scenario.}
In this scenario, two benchmarks share the same NUMA machine (for instance, hosted by a Cloud provider), as assumed in Section \ref{subsubsec:multiple-workloads}. %details. 
%This machine is governed by some co-scheduling run-time that assigns different partitions of the available computational nodes that each  application can use to run its threads. 
We assume that one benchmark, $A$, is not
memory-intensive and the other, $B$, is.
Hence, while $A$ will place pages locally for improved latency, $B$ will resort to \name to 
scatter its pages across all nodes (including the non-worker nodes where $A$ resides) in order to optimize $B$'s throughput, while not degrading $A$'s.

\begin{figure}[t!]
\begin{subfigure}{.45\textwidth}
  \centering
  \includegraphics[width=\textwidth]{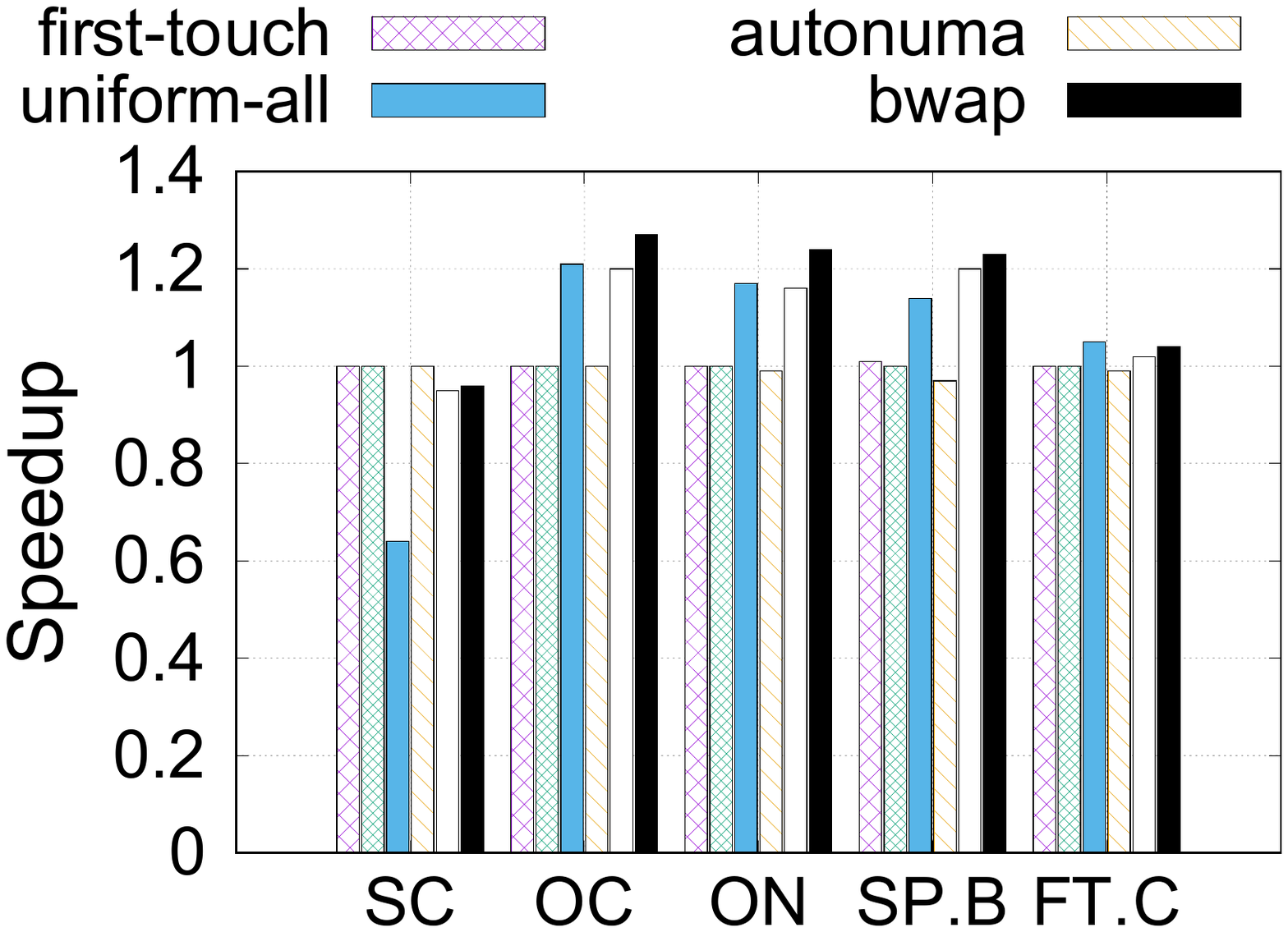}
  \caption{One worker node\\(co-scheduling, machine B)}
  \label{fig:machineBw1}
\end{subfigure}%
\hspace*{-0,4cm}
\begin{subfigure}{.45\textwidth}
  \centering
  \includegraphics[width=\textwidth]{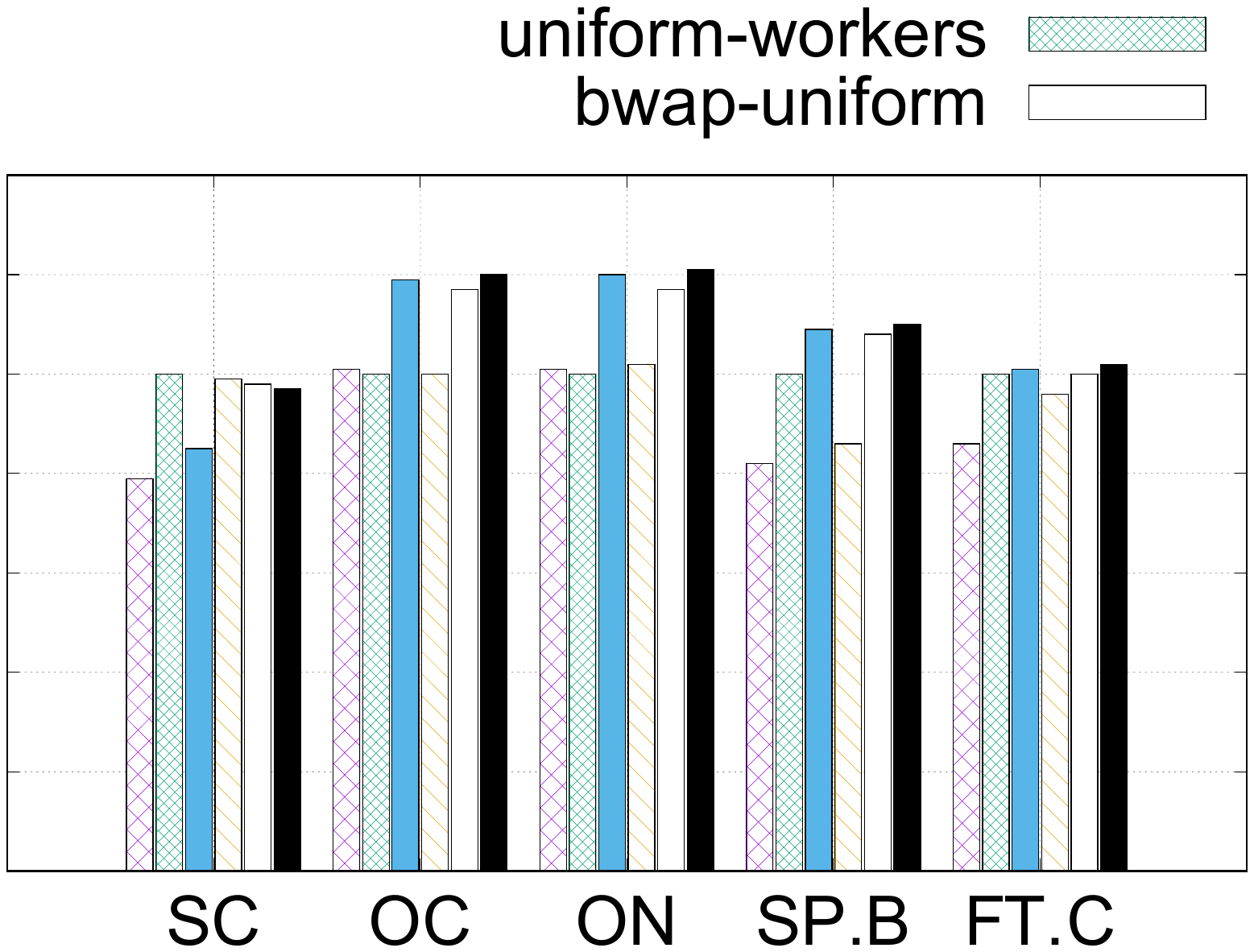}
  \caption{Two worker nodes\\(co-scheduling, machine B)}
  \label{fig:machineBw2}
\end{subfigure}\\
\begin{subfigure}{.45\textwidth}
  \centering
  \includegraphics[width=\textwidth]{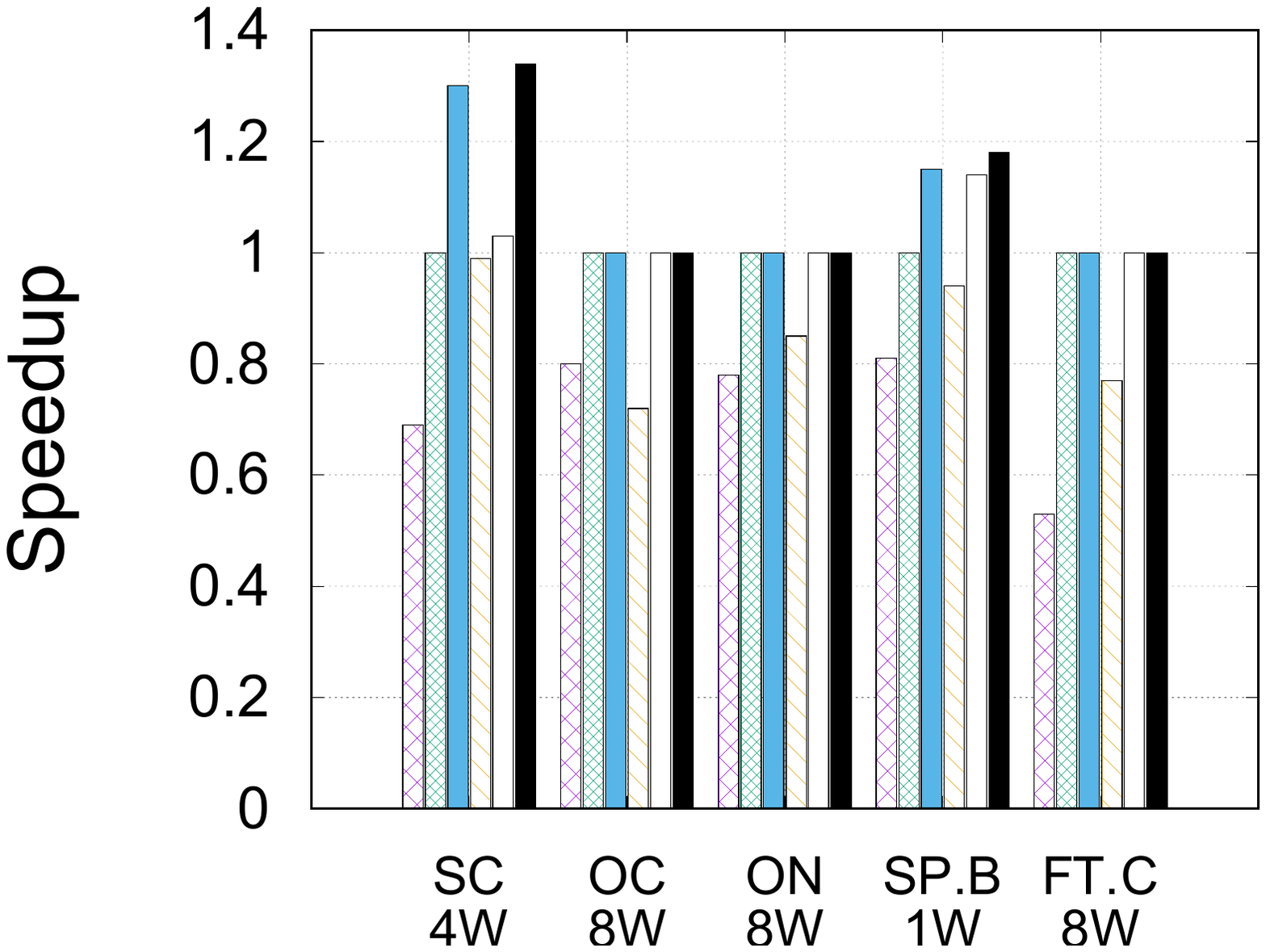}
  \caption{Optimal num.~of workers \\(stand-alone, machine A)}
  \label{fig:standaloneA}
  \end{subfigure}
  \hspace*{-0,4cm}
  \begin{subfigure}{.45\textwidth}
  \centering
  \includegraphics[width=\textwidth]{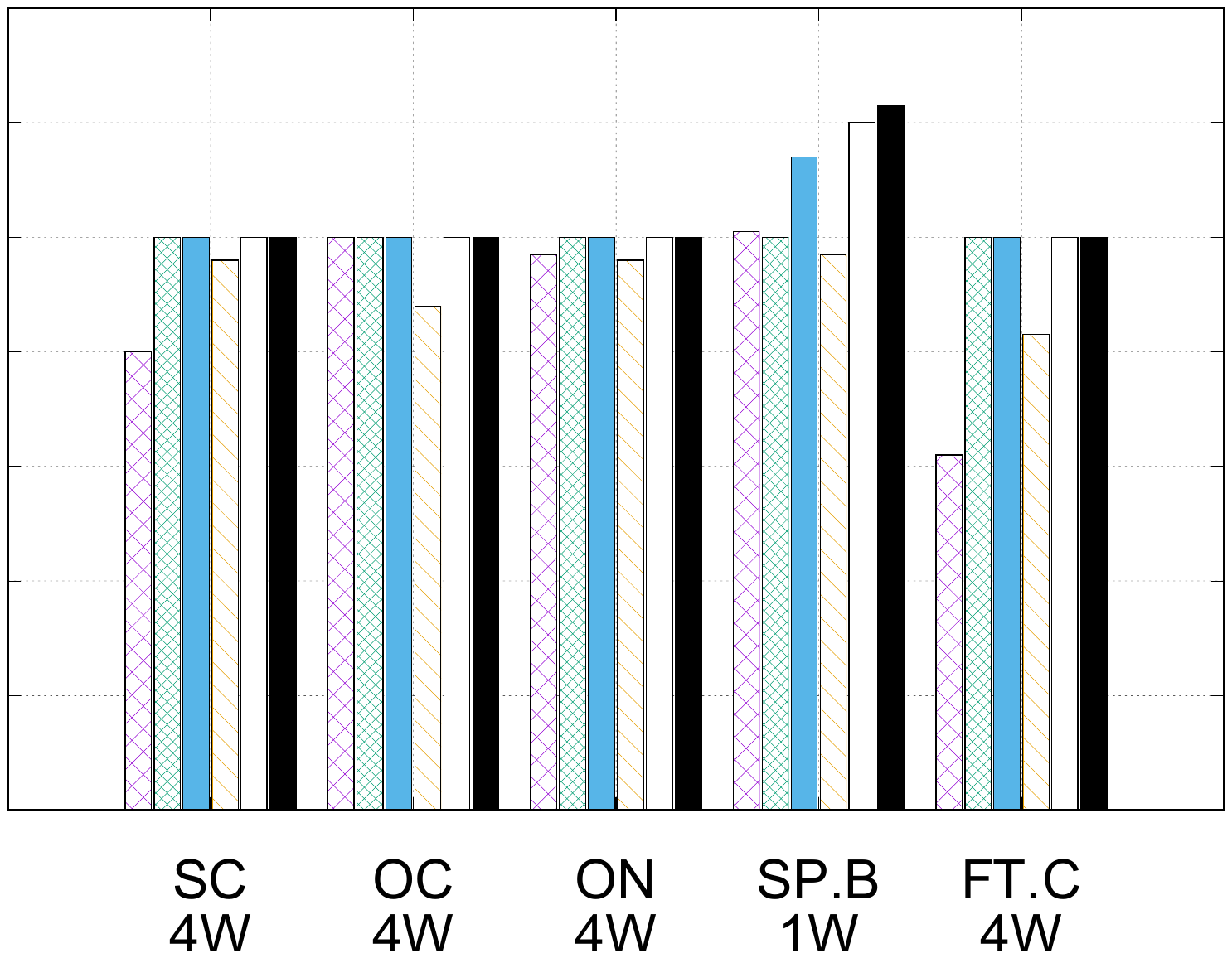}
  \caption{Optimal num.~of workers \\(stand-alone, machine B)}
  \label{fig:standaloneB}
\end{subfigure}%
\caption{Speedup vs \emph{uniform workers}} 
\label{fig:detailedmachineB}
\squeezeup
\end{figure}

Figures \ref{fig:detailedmachineA}, \ref{fig:machineBw2} and \ref{fig:machineBw1} compare the performance of each application $B$ on the co-scheduled scenario.
We consider different allocations of worker nodes to application $B$: $1$, $2$, and $4$ worker node(s) in machine A;  $1$, and $2$ worker node(s) in machine B. In all cases, $A$ runs on the remaining nodes.
As for application $A$, we run Swaptions \cite{parsec}. In all our co-scheduled experiments, we did not observe relevant changes to the performance of Swaptions when application $B$ placed some of its pages on the nodes of $A$, so we omit the performance results of this application in this analysis.

%Furthermore, we assume that the remaining node(s) run CPU-intensive applications that do not place relevant demand on the memory system. Refer to Section \ref{subsubsec:multiple-workloads} for the implementations details. 

\textbf{Stand-alone scenario.} In this %the stand-alone 
scenario, the NUMA machine is entirely available for application $A$, which can be deployed with any number of threads/workers. Hence, a rational user will run the application with its optimal parallelism level (assumed to be tuned \emph{a priori}). Note that, for each application, %the corresponding 
its optimal parallelism level may differ depending on the page placement policy. 
Figures \ref{fig:standaloneA} and \ref{fig:standaloneB} show how each application performs when running with the different page placement alternatives in this scenario.
%For now, let us skip the analysis of the \emph{\name-uniform} variant, whose performance 
%is discussed in detail in the next section. %the next section discuss in detailed. 
The performance of the \emph{\name-uniform} variant is discussed %in detail 
in Section \ref{sec:evaldetailed}.

\textbf{Analysis of the results}.
The results for both scenarios confirm that, as expected, the best performing solutions %for our selection of memory-intensive applications 
are those that fully exploit the available memory BW by placing shared pages across all nodes (\uniformall and \name), rather than restricting placement within the boundaries of the worker node set (\uniformworkers and \emph{autonuma}). 
%Naturally, the performance gap between these two kinds of approaches converges to null as the worker node set grows to the entire set of nodes in the system (i.e., non-workers tend to an empty set). 

%Not surprisingly
Unsurprisingly, \emph{first-touch} is usually the worst alternative for multi-worker scenarios by substantial margins, since it tends to centralize many shared pages on a single node (where the initializing threads run) as studied before~\cite{carrefour,shoal,smart_arrays}. Our results show that, even in a single worker scenario, binding the entire application's memory locally to the only worker node is suboptimal for memory-intensive applications.
Among the solutions that spread pages across all nodes, \name %consistently 
achieves the best performance or, with less favourable applications, performs comparably to the best solution (\uniformall). More precisely, \name is able to outperform both \uniformworkers and \emph{autonuma} by up to $1.66\times$, and \uniformall by up to $1.50\times$. 
As expected, the largest speedups of \name are observed on machine A, which has the most asymmetric topology. This confirms our main claim that, for some types of workloads, trivial uniform interleaving policies are not appropriate to exploit the full BW of complex NUMA systems.

One important trend is that %Figure \ref{fig:detailedmachineA}, \ref{fig:machineBw1} and \ref{fig:machineBw2} suggests that 
the benefits of \name over the uniform interleaving alternatives drop when more workers are involved. This is evident in the co-scheduled scenarios (as the number of workers increases). Further, in the stand-alone scenario, this explains why \name only achieves relevant gains for those applications whose optimal parallelism level is lower than the total number of nodes.
This trend can be explained by two factors. First, as  applications use an increasingly larger worker node set, the worker vs. non-worker dichotomy fades away. %In the \emph{all-workers} extreme, this dichotomy disappears. 
Thus, the gains from tuning \DWP decrease. % with larger worker node sets.
% However, even in an all-workers configuration, one may still question why \name's weighted interleaving among the worker set (i.e., independently of \name) is no longer advantageous. 
A second, %yet 
less intuitive factor, is that, as one enlarges the worker node set, the inter-worker canonical weight distributions (as devised by our \bwapA) tend to uniformity.

%\hl{
Let us now focus on the subset of benchmarks with non-negligible thread-private memory BW demand, namely OC, ON and FT.C.
For such benchmarks, consider the set of thread-private pages that belong to the threads running in a given worker node. In theory, the optimal placement of such pages should consider that worker node only (and regard every other node as non-worker).
However, both components of \mbox{\name} are, by design, based on the simplifying assumption that every page is accessed from \emph{every} worker node. This approximation allows \mbox{\name} to decide the placement of every page similarly, independently of each page's actual worker node set (i.e., the full worker node set in the case of shared pages; a specific worker node in the case of thread-private pages).

Despite this approximation, \name is still able to obtain important
performance gains with those benchmarks  
whose ratios of thread-private accesses are at odds with the above assumption (OC, ON and FT.C). 
The performance benefits obtained with these benchmarks follow the same trend as discussed above: the performance advantage of \name over the remaining alternatives is especially evident when the set of worker nodes is smaller; it decays as we enlarge the worker %node
set, becoming comparable to the best-performing alternative when worker nodes span across %every 
all nodes in the machine.

A more careful analysis allows us %to draw 
to shed some light on these results.
We start by observing that the memory demand that these benchmarks place on thread-private pages is sufficiently high to saturate the local memory BW. We support this observation by the fact that, when we place thread-private pages exclusively locally to the thread that accesses each page (with the \mbox{\emph{first-touch}} policy), their performance degrades relatively to the opposite extreme of \mbox{\uniformall}.
Therefore, the thread-private access demand of these benchmarks calls for page placement strategies that interleave pages across multiple nodes to maximize the available thread-private BW.

%On top of 
Based on this observation, we note that, among the evaluated strategies that resort to page interleaving to meet the above %previous 
requirement, \mbox{\name}'s \mbox{\emph{best-effort}} approach is the most effective one in scenarios where the %set 
number of worker nodes is small -- %namely, 
either one or two %nodes 
in our experiments. 
%More precisely, 
Specifically, when there is only a single worker node, \mbox{\name} is trivially accurate for thread-private pages. However, when there are two worker nodes, then \mbox{\name} will wrongly decide to place some thread-private pages in the nearest node instead of the local node (%when compared 
unlike to an optimal placement that takes %took 
thread-private pages into account).
Still, our results suggest that the impact of such an inherent inaccuracy is modest. This can be explained by the fact that the BW ratio between the local node and the nearest remote node is much lower ($1.7\times$ in machine A, $1.8\times$ in machine B) than the BW ratio between the local node and the farthest nodes ($5.1\times$ in machine A, $2.3\times$ in machine B). Consequently, despite its inaccurate modelling of thread-private accesses, \mbox{\name} still %manages to 
outperforms  alternative approaches whose placement decisions lead to even larger inaccuracies: those that rely on uniform interleaving do not take BW heterogeneity into account at all (thus it places high number of pages in the farthest nodes); while those that simply place thread-private pages locally fail to exploit the additional BW of the non-worker nodes.
Overall, these results suggest that \mbox{\name}'s approximated approach is 
accurate enough to be advantageous with a wide set of memory-intensive applications. 
%*******************************

\subsection{Detailed analysis of \name's components}
\label{sec:evaldetailed}

We study the gains and overheads of \name components.

\textbf{Canonical tuner.} 
Looking %back 
at Figures \ref{fig:detailedmachineA} and \ref{fig:detailedmachineB}, we can compare the performance of the full \name with \emph{\name-uniform}. 
The difference between these alternatives denote the effective contribution of \bwapA. %that the \bwapA has.
Our results show that, for most cases where \name %is advantageous over 
outperforms \uniformworkers and \uniformall, the larger speedup slice is due to the \bwapA (which is evident by observing how the results of \emph{\name-uniform} and \name differ). 
Concretely, enabling the \bwapA 
attains speedups of up to $1.32\times$ relatively to the uniform variant of \name. Further, the highest relative gains happen in machine A, which can be explained by its highest asymmetry.
In contrast, the simpler and less asymmetric topology of machine B makes \name essentially perform similarly to \emph{\name-uniform}. %; the relatively homogeneous weights in Table \ref{table:appconfigurations} confirm this.
This highlights that, in machines with pronounced BW asymmetries, simple approaches based on uniform interleaving or 2-level weighted interleaving (like \emph{\name-uniform}) are substantially suboptimal; however, in more symmetrical machines, simpler solutions may be an acceptable choice.%}

\textbf{\bwapB.}
Continuing the previous analysis, we can compare \uniformall with \emph{\name-uniform} to infer the actual benefits that the \bwapB %of \name 
contributes %attains by itself 
(i.e., departing from \uniformall rather than from the canonical distribution). 
Figures \ref{fig:detailedmachineA} and \ref{fig:detailedmachineB}  show that, while \emph{\name-uniform} outperforms \uniformall in many scenarios, that is not always the case. This can be explained by taking into account two considerations.
%We explain this by two main factors. 
First, for some scenarios, the optimal \DWP is $0$. In this case, \emph{\name-uniform} produces the same interleaving as \uniformall, but with the overhead of the online iterative search. Table \ref{tbl:scenarios} details the optimal \DWP for each application in each co-scheduled scenario. This table yields a high correlation between the null \DWP cases and the cases where \emph{\name-uniform} does not outperform \uniformall.
On the positive side, Table \ref{tbl:scenarios} also shows many cases where adapting \DWP allowed \emph{\name-uniform} to choose intermediate values that clearly outperform \uniformall. 
This results in performance gains of up to $1.49\times$ relatively to \uniformall (which are further increased if we consider the combination of both \name components).%} %of \name).

\begin{table}[]
\resizebox{\textwidth}{!}{
\begin{tabular}{|r|r|r|r|r|r|}
\hline
 & \multicolumn{3}{l|}{Machine A} & \multicolumn{2}{l|}{Machine B} \\ \hline
Application & 1 Worker & 2 Workers & 4 Workers & 1 Worker & 2 Workers \\ \hline
SC & 48.00\% & 0\% & 23.80\% & 100\% & 100\% \\ \hline
OC & 14.10\% & 0\% & 0\% & 0\% & 0\% \\ \hline
ON & 14.10\% & 16\% & 0\% & 0\% & 0\% \\ \hline
SP.B & 0\% & 0\% & 0\% & 15.20\% & 22.20\% \\ \hline
FT.C & 0\% & 16.30\% & 0\% & 30.30\% & 0\% \\ \hline
\end{tabular}
\caption{Ideal number of worker nodes and DWP values computed via \name  iterative search  (Co-scheduled scenario).}
\label{tbl:scenarios}}
\squeezeup
\end{table}

\textbf{Overhead and accuracy of \bwapB.}
Finally, we study the accuracy and overheads of the \bwapB.
To evaluate both aspects, we have manually deployed each application (at a given machine and scenario) with different \emph{static} values of \DWP and measure the corresponding execution times and average stall rates. This allows us to understand, what the optimal \DWP is, what is the corresponding (maximum) performance, and the stall
rate curve effectively guides us towards finding the optimal \DWP. %Moreover, by 
By comparing with the value that our \bwapB chooses and the resulting execution time, we can assess how close to optimal our search gets and at which overhead.

\begin{figure}[!t]
\centering
\includegraphics[width=0.48\textwidth]{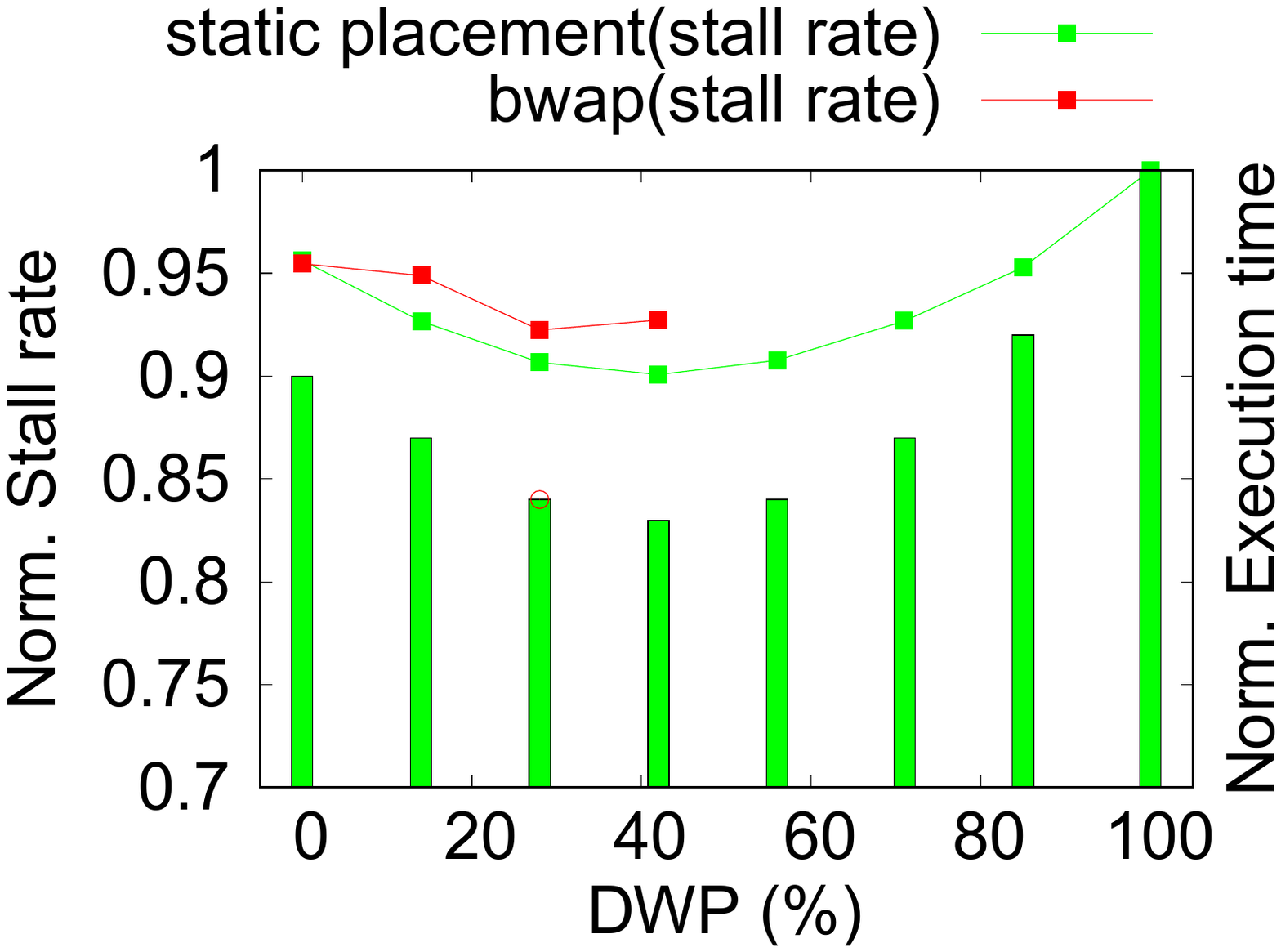}
\hspace*{-0,4cm}
\includegraphics[width=0.48\textwidth]{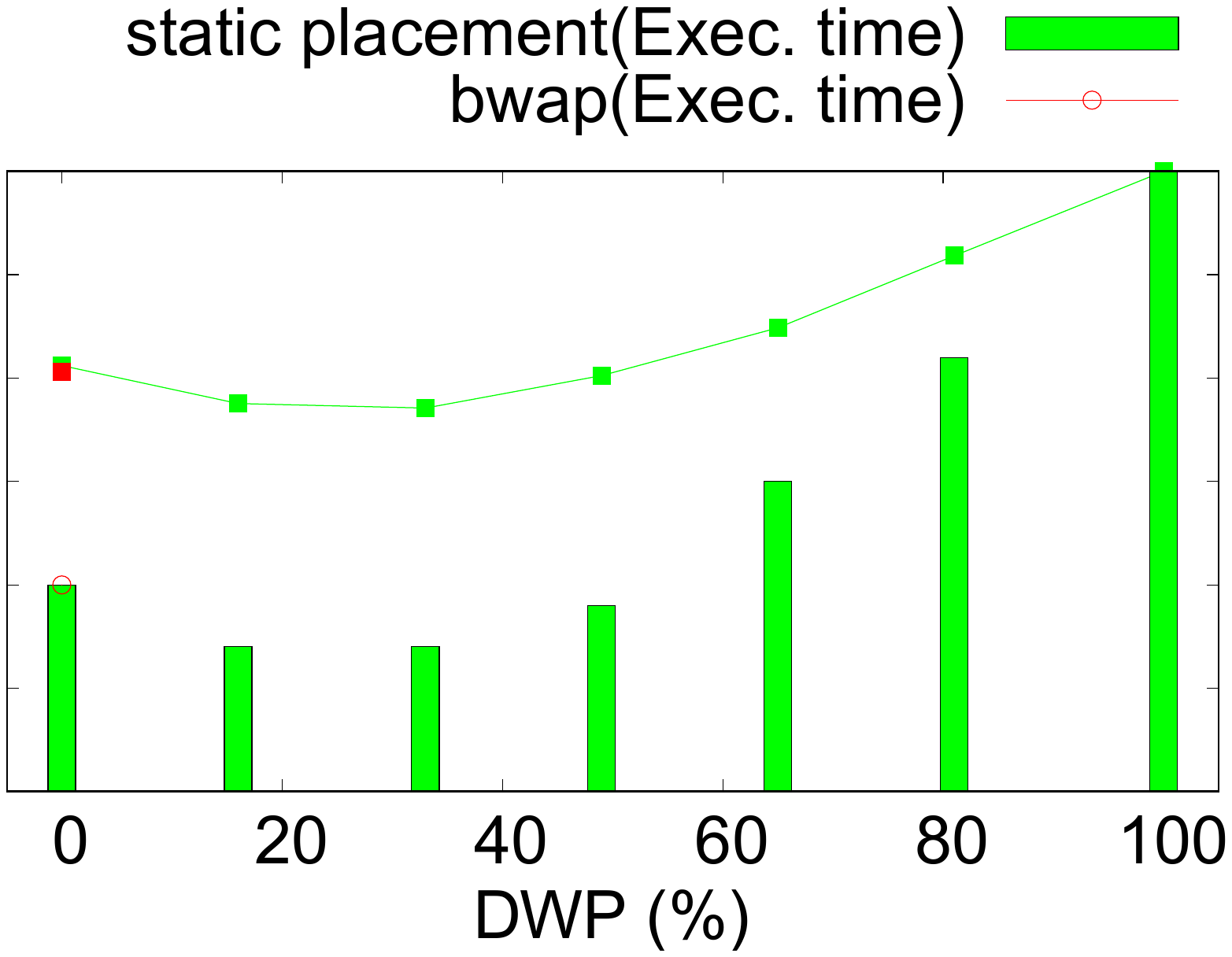}
\caption{Evaluating the \DWP iterative search using Streamcluster on machine A, with 1 (left) and 2 worker nodes (right).}
\label{fig:q1}
\squeezeup
\end{figure}

%\hl{
Our complete experiments with the entire selection of applications confirmed that stall rate is effectively correlated to execution time and its variation with \DWP is essentially convex. Furthermore, the \bwapB was able to successfully find the optimal \DWP by a maximum error margin of 1 iterative step for all cases we evaluated. Figure \ref{fig:q1} illustrates this with the Streamcluster application on machine A.

%sell the good news first
Regarding execution overhead, in the %set of 
experiments discussed in Section
\ref{sec:evaluation:apps}, we measured a maximum overhead of $4\%$ over all the applications. We found that the overhead of \bwapB can be increased by two main factors: i) higher optimal values of \DWP, since the search starts at the opposite extreme and each iteration costs time and page migrations; and ii) lower %total 
execution times, which do
not allow to amortize the cost of the search %performed by the \bwapB 
and benefit from resulting optimized page placement.

\section{Related Work}
\label{sec:relwork}

When deploying an application on a NUMA system, a number of complex questions arise, from how many threads, to where to place threads and pages. As NUMA systems increase their prominence, these problems receive increasing attention from the research community. %
% I propose we relate only to works that also address page placement
Optimizing thread and memory placement on NUMA systems has been extensively studied \cite{memprof,lira,asymsched,memorycontroller,carrefour,memory-access-pattern,memory-system,Pusukuri:2015:TEL:2836331.2827698,Srikanthan:2016:CSL:3026959.3026989,nucore,Harris:2015:CFP:2813767.2813771,pandia,kernel-based,Majo:2017:LPC:3057718.3040222,Georgakoudis:2017:SSP:3154814.3158643}.

Linux provides %several 
extensions\cite{autonuma,Sched-NUMA} to improve data access locality in NUMA systems. However, these extensions 
%do not take into account 
do not account for the interconnect asymmetry and do not improve %the 
BW for communicating threads. For in\-stance, autonuma\cite{autonuma} implements locality-driven optimization by migrating threads closer to the memory they access, and/or by migrating data to the memory closer to the threads.
%migrates threads closer to the memory they are accessing and/or migrates application data to memory closer to the threads that reference it, thereby implementing locality-driven optimization. 
Linux also provides an option to uniformly interleave part of address space across memory nodes. \name complements these strategies by providing a novel BW-aware alternative to the %trivial 
uniform interleaving.

 Considering proposals for page placement, we distinguish proposals aiming at minimizing memory latency (e.g., \cite{kernel-based,autonuma}) and proposals 
 aiming at optimizing memory BW %that consider memory BW as the main bottleneck 
 (e.g., \cite{asymsched,carrefour}).
Depending on the application, the main bottleneck can be latency or BW, or somewhere between those extremes. 
In contrast to the proposals that focus on 
either of extremes, \name takes this spectrum into account with its \bwapB.

Apart from page placement, other approaches have been proposed to improve memory %access
performance in NU\-MA systems. 
Carrefour \cite{carrefour} replicates 
read-only pages accessed from multiple nodes. %pages that are read-only and accessed from multiple nodes. In addition,
It also detects private pages and places them close to the corresponding thread. Shoal \cite{shoal} and Smart Arrays \cite{smart_arrays} introduce programming abstractions that enable a runtime layer to choose the most appropriate %NUMA 
page placement, data layout or replication, while taking into account the underlying memory topology and the run-time behaviour of the application. Shoal and Smart Arrays adopt uniform interleaving as one of their NUMA-aware page placements. \name is less intrusive, since the application only needs to be linked with and activate \bwapB. Still, the design principles behind \name's placement policy can also be applied to improve %the placement and replication decisions in 
these systems.

Recently, some authors have studied the problem of BW-aware page placement for different heterogeneous memory systems \cite{batman,gpu,BA-ics}. 
All these works consider a tiered memory architecture consisting of two or more types of memory nodes with heterogeneous performance characteristics (BW, latency, etc).
Moreover, all works share the fundamental insight that, 
%for BW-intensive applications, their throughput
performance of BW-intensive applications improves if memory accesses are distributed across all memories, proportionally to the BW of each memory. 
%Multi-node 
NUMA systems, as addressed by \name, are also characterized by heterogeneous BWs, even when every NUMA node relies on the same memory technology (e.g., DRAM). Hence, \name shares the same BW-aware placement principle with the previous works. 
However, the problem of BW-aware page placement in NUMA systems has crucial differences (as Section \ref{sec:introduction} highlights), with new challenges that render the above-mentioned proposals either inappropriate or strongly suboptimal %if applied directly to
for NUMA systems.
%Among Some of 
The aforementioned %previously mentioned 
proposals for BW-aware page placement %in heterogeneous memory systems they 
either do not support %multi-node 
NUMA systems \cite{batman} or use the \texttt{uniform-workers} policy to distribute pages across (hybrid memory) NUMA nodes.
Hence, \name can complement or extend these techniques to support NUMA systems whose nodes comprise heterogeneous hybrid memory hierarchies.

Our work is also related to workload consolidation techniques for cloud computing systems and datacenters. %, where the co-scheduled scenarios that \name supports are the norm. 
Recent works have focused on optimizing memory throughput in these scenarios by employing last-level cache and memory band-width partitioning \cite{copart,emba} across co-scheduled applications. While these works focus on single-node systems where two or more applications share the same CPU, \name targets multi-node NUMA systems where each node is exclusively allocated to one application at most.
Works such as \cite{kaestle,mctop} propose effective tools to characterize (either through an analytical model or through an empirical procedure) the NUMA topology. These can be integrated into \name to allow \name to devise a more accurate canonical distribution.

\section{Conclusions}
\label{sec:conclusions}
%\improvement[inline]{REVISION NEEDED}
Although new thread placement approaches for asymmetric NUMA systems have recently emerged, today's usual techniques for page placement still rely on the obsolete assumption of a symmetric architecture.
This paper proposes \name, a novel approach for asymmetric %bandwidth
BW-aware placement of %shared 
pages in NUMA systems. 
Our %experimental 
evaluation shows that \name improves the gains of state-of-the-art policies by up to $66\%$, on commodity NUMA machines. The gains of \name are especially evident in co-scheduled scenarios and when the application does not scale up to the available hardware parallelism.%, and/or in co-scheduled environments where the memory-intensive application is only allowed to use a subset of CPUs of a larger NUMA machine.

%\hl{
While far from trivial, the design of \mbox{\name} is inherently \emph{best-effort}, since it relies on important simplifying assumptions about the workloads and the underlying system. 
Therefore, our contributions can be seen as first step that opens avenues for follow-up research on BW-aware page placement for NUMA systems.
Among future work directions, we plan to more accurately model workloads with relevant write and/or thread-private access volumes, as well as workloads with non-uniform access distributions to the shared address space. 
Since these scenarios are characterized by inherently asymmetric memory access patterns,
the different sets of pages may have distinct optimal placements (e.g., depending on whether a page is thread-private or shared, read or write-dominated, hot or cold). 
Hence, accurately determining the optimal page placement in such scenarios requires devising different canonical weight distributions and \mbox{\DWP} values, as well as physically mapping pages according to different placement configurations. The key challenge here lies in achieving these goals while retaining
%Achieving this while retaining 
\mbox{\name}'s key virtues of %practicality 
transparency and portability. %is intrinsically hard.
% Therefore, every component of \mbox{\name} would need to handle different page ranges differently, according to their specific access patterns.
% . The necessary changes to \mbox{\name} will 
%  As an example, let us consider the changes that we would need to introduce to accurately take thread-private pages into consideration in \mbox{\name}. 
% Regarding the \mbox{\bwapA}, different canonical weight distributions would need to be calculated for the thread-private pages of every other worker node. Likewise, the \mbox{\bwapB} would no longer have to perform a one-dimensional search, since the \mbox{\DWP} of different page ranges would now be optimized against distinct reference worker sets.
% %For those applications, thread-private pages should also be interleaved across multiple memory nodes, according to adequate weights, in order to benefit from higher BW. 
% %These deep changes pose strong obstacles to a practical implementation. 
% Besides the need to collect richer information about the memory access behaviour of the application, the canonical distribution would become a function of application-specific metrics (thus, no longer being application-agnostic). %Further, the \mbox{\bwapB} would no longer be a one-dimensional search.
% Finally, it would be harder to enforce the resulting page placement, as the input would now be a set of per-page range weight configurations.

As further future work, we intend to extend \mbox{\name} to dynamically adjust its weight distribution throughout the application's execution time, in order to obtain improved performance for applications whose access patterns change over time or for co-scheduling scenarios with dynamic sets of applications. This extension would enable integrating \mbox{\name} at the core of dynamic runtime support systems like Callisto \mbox{\cite{callisto}} or Asymsched \mbox{\cite{asymsched}}. Finally, we plan to extend \mbox{\name} to support NUMA systems whose nodes have hybrid memory subsystems (e.g. DRAM and NVRAM) or where the computational power is heterogeneous between different nodes.

\section{Acknowledgements}
%The research leading to these results 
This work has received funding from the 
European Union’s Horizon 2020 research and innovation programme under the EPEEC project, grant agreement No 801051;
% %European Union’s 
% EU Horizon2020 %research and innovation 
% programme %under 
% via the EPEEC project %(Grant Agreement %No
% (GA 801051); 
from the %Foundation for Science and Technology 
Fundação para a Ciência e a Tecnologia %(FCT) 
via the projects UIDB/50021/2020, and PTDC/EEI-SCR/1743/2014; and from the Erasmus Mundus Joint Doctorate in Distributed Computing (EMJD-DC) funded by the Education, Audiovisual and Culture Executive Agency %(EACEA) 
of the European Commission under FPA 2012-0030. We thank Amin Mohtasham for sharing his initial findings on the behaviour of BW-intensive applications in NUMA, which inspired us to pursue this work.

%%%%%%% -- PAPER CONTENT ENDS -- %%%%%%%%

%%%%%%%%% -- BIB STYLE AND FILE -- %%%%%%%%
%\input{ref_shorten.tex}

\bibliographystyle{plain}
\bibliography{ref_b}
%\bibliography{ref}

%%%%%%%%%%%%%%%%%%%%%%%%%%%%%%%%%%%%

\end{document}